\documentclass[10pt,ams]{article}
\usepackage{graphicx}
\usepackage{amssymb}
\usepackage{amsmath}

\newcommand{\p}{\partial}

\newcommand{\s}{\sigma}

\renewcommand{\d}{\delta}
\renewcommand{\S}{\Sigma}
\newcommand{\Int}{\int{d^{4}\!x}}

\newcommand{\bci}{\bar{c}^{i}}

\newcommand{\half}{\mbox{$\frac{1}{2}$}}
\begin{document}
\date{}
\title{\textbf{Study of the nonlocal gauge invariant mass operator
 $\mathrm{Tr} \int d^4x
F_{\mu\nu} (D^2)^{-1} F_{\mu\nu}$ in the maximal Abelian gauge}}
\author{ \textbf{M.A.L. Capri$^a$\thanks{marcio@dft.if.uerj.br}}\,,
\textbf{V.E.R. Lemes$^{a}$\thanks{vitor@dft.if.uerj.br}}\,,
 \\\textbf{R.F. Sobreiro}$^{b}$\thanks{%
sobreiro@cbpf.br}\,, \textbf{S.P. Sorella}$^{a}$\thanks{%
sorella@uerj.br}{\ }{\ }\footnote{Work supported by FAPERJ,
Funda{\c c}{\~a}o de Amparo {\`a} Pesquisa do Estado do Rio de
Janeiro, under the program {\it Cientista do Nosso Estado},
E-26/100.615/2007.}\;, \textbf{R.
Thibes$^{a}$\thanks{thibes@dft.if.uerj.br}} \\\\
\textit{$^{a}$\small{UERJ
$-$ Universidade do Estado do Rio de Janeiro,}}\\
\textit{\small{Instituto de F\'{\i }sica, Departamento de F\'{\i
 }sica Te\'{o}rica,}}\\
\textit{\small{Rua S{\~a}o Francisco Xavier 524, 20550-013
Maracan{\~a}, Rio de Janeiro, Brasil}} \\ [2mm]
\textit{$^{b}$\small{CBPF $-$ Centro Brasileiro de Pesquisas
 F\'{\i}sicas,}} \\
\textit{\small{Rua Xavier Sigaud 150, 22290-180, Urca, Rio de
Janeiro,
 Brasil}}\\ } \maketitle
\begin{abstract}
The nonlocal gauge invariant mass operator $\mathrm{Tr} \int
d^{4}\!x\, F_{\mu\nu} (D^2)^{-1} F_{\mu\nu}$ is investigated in
Yang-Mills theories in the maximal Abelian gauge. By means of the
introduction of auxiliary fields a local action is achieved,
enabling us to use the algebraic renormalization in order to prove
the renormalizability of the resulting local model to all orders
of perturbation theory.
\end{abstract}
%\newpage
%\tableofcontents
%\newpage
%%%%%%%%%%%%%%%%%%%%%%%%%
\section{Introduction}

One of the major open problems in quantum field theory is the
understanding of nonabelian gauge theories, and consequently of
quantum chromodynamics (QCD), in the infrared regime. The
confinement phenomenon of quarks and gluons is not yet clearly
established from the theoretical point of view and still waits for
a satisfactory explanation.\\\\The Yang-Mills (YM) theories are
described by the following Euclidean action
\begin{equation}
 S_{\mathrm{YM}} = {1 \over 4} \int d^{4}\!x\, F^A_{\mu\nu} F^A_{\mu\nu}
\label{YM} \;,
\end{equation}
where $ F^A_{\mu\nu} $ is the field strength
% written in terms of the gauge vector potential $A_\mu$ as
\begin{equation}
 F^A_{\mu\nu} = \partial_\mu A_\nu^A - \partial_\nu A_\mu^A + g
 f^{ABC}A_\mu^B A_\nu^C
\label{Fmunu} \;.
\end{equation}
Here $f^{ABC}$ are the structure constants of the gauge group
$SU(N)$ with $A = 1,\dots,N^2-1$, and $g$ is the coupling
constant. At high energies, the running coupling constant is
sufficiently small to allow for a perturbative description, as
expressed by the asymptotic freedom
\cite{GrossWilczek:1973,Politzer:1973fx}. However, when one lowers
the energy, the running coupling constant grows, causing
perturbation theory to fail, so that nonperturbative techniques
are required. \\\\To deal with this problem, different approaches
have been considered. For example, in the Landau gauge, several
analytic results have been obtained on the infrared behavior of
the propagators of the theory, see for instance
\cite{Gribov:1977wm,Zwanziger:1989mf,Zwanziger:1992qr,von
 Smekal:1997is,von
 Smekal:1997vx,Alkofer:2000wg,Alkofer:2004it,Fischer:2006vf,Pawlowski:2003hq}.
In this gauge, lattice simulations have confirmed an infrared
suppressed gluon propagator exhibiting violation of positivity
\cite{Cucchieri:2004mf,Bowman:2007du,Cucchieri:2007md,Bogolubsky:2007ud,Furui:2006rx},
a feature interpreted as a signal of confinement. In particular,
the fitting of the lattice data for the gluon propagator are
usually accomplished with the aid of several mass parameters
\cite{Cucchieri:2004mf,Bloch:2003sk,Cucchieri:1997dx}, whose
theoretical interpretation is still under investigation. So far,
two possible origins could be suggested for such massive
parameters, namely: the existence of the Gribov copies
\cite{Gribov:1977wm} and the condensation of suitable dimension
two operators built up with gluon and ghost fields
\cite{Gubarev:2000eu,Gubarev:2000nz,Verschelde:2001ia}. In this
work we shall focus on dimension two operators built up with gluon
fields only, see \cite{Capri:2005vw,Capri:2007hw} for a recent
discussion of ghost condensation in the Landau and maximal Abelian
gauges.
\\\\The massive Gribov parameter $\gamma$, which is fixed by a gap
equation, follows from the restriction of the domain of
integration in the Feynman path integral to the so called Gribov
region \cite{Gribov:1977wm,Zwanziger:1989mf,Zwanziger:1992qr}.
This restriction is needed in order to deal with the problem of
the Gribov copies and may be implemented through the introduction
in the YM action of a nonlocal term, known as the Zwanziger
horizon function \cite{Zwanziger:1989mf}. However, there is still
room for additional mass parameters. Therefore, the possibility of
the condensation of dimension two operators, giving rise to a
dynamically generated mass for the gluons, has been taken into
account. Let us also mention that, besides the lattice data, the
introduction of an effective gluon mass turns out to be useful
also from the phenomenological point of view, see for example
\cite{Parisi:1980jy,Field:2001iu,Giacosa:2004ug}. In particular,
in the Landau gauge, the dimension two operator $A^A_\mu A^A_\mu$
was proven to be multiplicatively renormalizable to all orders
\cite{Dudal:2002pq}. In
\cite{Verschelde:2001ia,Browne:2003uv,Dudal:2003vv,Sorella:2006ax},
an effective potential for $A^A_\mu A^A_\mu$ was constructed. The
formation of a nonvanishing condensate $\langle A^A_\mu A^A_\mu
\rangle $, resulting in a dynamical effective gluon mass, turned
out to be energetically favored.
\\\\Besides the Landau gauge, other gauge fixings have been
considered. We mention the recent analysis of dimension two
operators in the Curci-Ferrari \cite{Dudal:2003gu} and general
linear covariant gauges \cite{Dudal:2003by,Dudal:2003np}. In
\cite{Dudal:2003gu}, an effective potential was constructed for
the on-shell BRST invariant operator $(\half A^A_\mu A^A_\mu +
\alpha {\bar c}^A {\bar c}^A)$ in the Curci-Ferrari gauge while,
in \cite{Dudal:2003by,Dudal:2003np}, a detailed study of the
already mentioned operator $A^A_\mu A^A_\mu$ was performed in the
linear covariant gauges. As a result, in \cite{Dudal:2003np}, it
was shown that the gluons do acquire a dynamical mass since the
formation of $\langle A^A_\mu A^A_\mu \rangle$ is energetically
favored. Another interesting gauge which is receiving increasing
attention in the last years is the  maximal Abelian gauge (MAG)
\cite{'t
 Hooft:1981ht,Kronfeld:1987vd,Kronfeld:1987ri}.  Several
results have already been established in this gauge, both from
theoretical
\cite{Dudal:2004rx,Capri:2006cz,Capri:2005tj,Bruckmann:2000xd,Gracey:2005vu}
and lattice \cite{Amemiya:1998jz,Bornyakov:2003ee,Mendes:2006kc}
points of view.  The MAG is well suited for the study of special
aspects of infrared QCD and color confinement as, for instance,
the dual superconductivity and the so called Abelian dominance.
The dual superconductivity mechanism
\cite{Nambu:1974zg,thooft:75,Mandelstam:1974pi} asserts that the
low energy regime of Yang-Mills theories should exhibit monopoles
as vacuum configurations. The condensation of these magnetic
charges might give rise to a dual Meissner effect in the
chromoelectric sector.  As for the Abelian dominance hypothesis
\cite{Ezawa:1982bf}, the infrared limit of QCD should be described
by an effective theory constructed only from Abelian degrees of
freedom, identified with the diagonal components of the gauge
field, corresponding to the generators of the Cartan subgroup of
the gauge group.  Lattice numerical simulations in the MAG have
reported significant differences between the diagonal and and
off-diagonal components of the gluon propagator
\cite{Amemiya:1998jz,Bornyakov:2003ee,Mendes:2006kc}. In
particular, the off-diagonal gluon propagator displays a mass
greater than that reported for the diagonal component,
corroborating in fact the Abelian dominance hypothesis. In an
attempt to understand those lattice results in the MAG, also here
the condensation of dimension two operators has been considered.
In \cite{Dudal:2004rx,Dudal:2005bk}, a dynamical mass generation
mechanism for the off-diagonal gluons was proposed in the MAG, by
means of the condensation of the operator $\langle \half A^a_\mu
A^a_\mu + \alpha {\bar c}^a {\bar c}^a \rangle$\footnote{Here the
index $a$ runs only on the off-diagonal components, see the
beginning of the next section for the notations.}. \\\\As the
reader may have noticed, the dimension two operators mentioned
above are gauge dependent, being related to specific choices of
the gauge fixing. This is a consequence of the fact that a {\it
local gauge invariant} dimension two operator is not available in
YM theories. Still, the condensation of these operators might be
taken as evidence in favor of the existence of a more fundamental
{\it gauge invariant} operator. However, willing to preserve gauge
invariance, we are led to give up of the locality requirement. The
price one has to pay for that is that nonlocal operators are
difficult to be handled within a consistent renormalizable
framework. So far, several possibilities have been considered. The
first proposal for a condensate of dimension two was made by
\cite{Gubarev:2000eu,Gubarev:2000nz}, who considered the nonlocal
gauge invariant operator $A^2_{\rm min}$, obtained by minimizing
the operator $A^A_\mu A^A_\mu$ along the gauge orbit, namely
\begin{equation}
A^2_{\rm min}  =  \min_{\{U\}}  \Int\; Tr \left( A^{U}_\mu
\right)^2 \,, \label{A2min}
\end{equation}
where $U$ represents an element of the gauge group $SU(N)$,
$A^{U}_\mu = UA_{\mu}U^{-1} +i U\partial_{\mu}U^{-1}$. However,
for a generic choice of the gauge fixing, the operator $A^2_{\rm
min} $ proves to be very difficult to be handled at quantum level.
Expanding $A^2_{\rm min} $ in a power series in the gauge field,
see for example \cite{Capri:2005dy}, one obtains
\begin{equation}
 A^2_{\rm min} = \Int \left[ A^A_\mu
\left( \delta_{\mu\nu} - \frac{\partial_\mu
\partial_\nu}{\partial^2} \right)A^A_\nu -g f^{ABC}
\left(\frac{\partial_\nu}{\partial^2} \partial A^A \right) \left(
\frac{1}{\partial^2} \partial A^B \right) A^C_\nu \right]
+O(A^4)\;. \label{Oexp}
\end{equation}
The series \eqref{Oexp} consists of an infinite number of nonlocal
terms. So far, a consistent treatment for $A^2_{\rm min} $ has
been achieved only in the Landau gauge, $\partial_{\mu}
A^A_{\mu}=0$, where all the non-local terms vanish and
\eqref{Oexp} simplifies to the already mentioned operator $ \Int
\left( A^{A }_\mu \right)^2 $. Let us also quote here that, in a
generic linear covariant gauge, the anomalous dimension of
$A^2_{\rm min}   $ has been calculated at one loop order in
\cite{Gracey:2007ki}.
\\\\More recently, we have pointed out \cite{Capri:2007ck} that
another kind of gauge invariant nonlocal operator might be
relevant, namely
\begin{equation}
{\cal O} = \mathrm{Tr} \int d^4x F_{\mu\nu}{(D^2)}^{-1} F_{\mu\nu}
\label{trmassop} \;.
\end{equation}
Unlike expression \eqref{A2min}, the operator \eqref{trmassop} has
the advantage of being localizable by means of the introduction of
a suitable set of auxiliary fields. Firstly introduced in the case
of 3d YM \cite{Jackiw:1997jg}, the operator ${\cal O}$ has
received renewed interest in the context of 4d YM. In fact, we
have been able to show that, when cast in local form, it gives
rise to a local action which can be proven to be renormalizable to
all orders in the general class of the linear covariant gauges
\cite{Capri:2005dy}. Moreover, in \cite{Capri:2006ne}, the
anomalous dimension of \eqref{trmassop} has been evaluated at the
two loop order in the $\bar{MS}$ scheme and explicitly proven to
be independent from the gauge parameter. Also, in the case of the
Landau gauge, it has been shown \cite{Capri:2007ix} that the
inclusion of Zwanziger's horizon function does not spoil the
renormalizability of \eqref{trmassop}.
\\\\ Despite the progress already achieved in the Landau and covariant linear gauges,
a detailed analysis of the gauge invariant operator ${\cal O}$ in
the MAG is still lacking. The main task of the present paper is to
fill this gap, {\it i.e.} to achieve a local and renormalizable
framework for ${\cal O}$ in the MAG. In particular, the manifest
gauge invariance of the operator ${\cal O}$ might be useful in
order to improve our present understanding of issues like the
Abelian dominance and the dynamical gluon mass generation in the
MAG.
\\\\This paper is organized as follows.  In section {\bf 2}, by
means of the introduction of auxiliary fields, we obtain a local
action for YM theories in the MAG, in the presence of the mass
operator \eqref{trmassop}. Moreover, the embedding of the
resulting local theory in a more general action, enables us to
make use of the BRST transformations. In section {\bf 3} we obtain
the full set of Ward identities fulfilled by the starting action.
Section {\bf 4} is devoted to the proof of the renormalizability
of this action to all orders of perturbation theory.  We obtain
the most general invariant counterterm and we prove that it can be
reabsorbed by means of a redefinition of fields and parameters of
the starting action. The last section collects our conclusions.
Appendix {\bf A} contains a detailed discussion of the mass
operator \eqref{trmassop} in the presence of the horizon function
for the MAG.

%%%%%%%%%%%%%%%%%%%%%%%%%%%%
\section{Local action in the maximal Abelian gauge}

As is well known, the action (\ref{YM}) is left invariant by the
gauge transformations
\begin{equation}
 \delta_\omega A^A_\mu = - D^{AB}_\mu \omega^B \;,
\end{equation}
for arbitrary $\omega^A(x)$.  In order to quantize the theory, one
must fix the gauge.  As we shall choose an Abelian gauge, we
decompose the
 field ${\cal A}_\mu$ as
\begin{equation}
 {\cal A}_\mu = A^A_\mu T^A \equiv A^a_{\mu} T^a + A^i_\mu T^i
\;,
\end{equation}
with
\begin{eqnarray}
 \left[ T^a, T^b \right] &=& igf^{abc}T^c + igf^{abi}T^i \;,\\
 \left[ T^i, T^a \right] &=& igf^{iab}T^b \;,\\
 \left[ T^i, T^j \right] &=& 0 \;.
\end{eqnarray}
The index $ i = 1,\dots,N-1 $ labels the $N-1$ diagonal generators
 $\{T^i\}$ of the Cartan subalgebra of $SU(N)$, while the index $a =
 1,\dots,N(N-1)$ labels the remaining off-diagonal generators
 $\{T^a\}$. \\\\Accordingly, the field strength decomposes as
\begin{eqnarray}
 F^a_{\mu\nu} &=& D^{ab}_\mu A^b_\nu - D^{ab}_\nu A^b_\mu + gf^{abc}
 A^b_\mu A^c_\nu
 \;,
 \\
 F^i_{\mu\nu} &=& \partial_\mu A_\nu^i - \partial_\nu A_\mu^i +
 gf^{abi} A^a_\mu A^b_\nu
 \;,
\end{eqnarray}
where we have introduced the covariant derivative $D^{ab}_\mu$
with respect to the diagonal components $A^i_\mu$ of the gauge
field, namely
\begin{equation}
 D^{ab}_\mu \equiv \delta^{ab}\partial_\mu - gf^{abi} A^i_\mu
 \label{covdev}
\;.
\end{equation}

\subsection{Gauge fixing}
The MAG \cite{'t Hooft:1981ht,Kronfeld:1987vd,Kronfeld:1987ri} is
obtained by requiring that the functional
\begin{equation}
 {\cal R}[ A ] =  \int d^4 x \; A_\mu^a A^{a}_\mu
 \label{A2}
\end{equation}
is stationary with respect to the gauge transformations. Observe
that expression (\ref{A2}) depends only on the off-diagonal
components of the gauge field. The vanishing of the first
variation of ${\cal R}$ leads to the non-linear condition
\begin{equation}
 D^{ab}_\mu A_\mu^b = 0
 \label{MAGgf}
\;.
\end{equation}
Still, it remains to choose a gauge condition for the diagonal
components $A^i_\mu$ of the gauge field. We shall impose a Landau
type condition, also employed in lattice numerical simulations,
{\it i.e.}
\begin{equation}
\partial_\mu A^i_\mu = 0
 \label{Abeliangf}
\;.
\end{equation}
The conditions (\ref{MAGgf},\ref{Abeliangf}) are implemented by
adding the following gauge fixing term to the Yang-Mills action
\begin{eqnarray}
 S_{\mathrm{MAG}}&\!\!\!=\!\!\!&\int
d^{4}\!x\,\left(b^{a}D^{ab}_{\mu}A^{b}_{\mu}
+\bar{c}^{a}D^{ab}_{\mu}D^{bc}_{\mu}c^{c}
+gf^{abi}\bar{c}^{a}(D^{bc}_{\mu}A^{c}_{\mu})c^{i}
+\bar{c}^{a}D^{ab}_{\mu}(gf^{bcd}A^{c}_{\mu}c^{d})\right.\nonumber\\
&&\phantom{\int
d^{4}\!x\,}\left.-g^{2}f^{abi}f^{cdi}\bar{c}^{a}c^{d}A^{b}_{\mu}A^{c}_{\mu}
+b^{i}\partial_{\mu}A^{i}_{\mu}
+\bar{c}^{i}\partial_{\mu}(\partial_{\mu}c^{i}
+gf^{abi}A^{a}_{\mu}c^{b})\right)\nonumber\\
&&+\alpha\int d^{4}\!x\,\left(\frac{1}{2}\,b^{a}b^{a}
-gf^{abi}b^{a}\bar{c}^{b}c^{i}
-\frac{g}{2}f^{abc}b^{a}\bar{c}^{b}c^{c}
-\frac{g^{2}}{4}f^{abi}f^{cdi}\bar{c}^{a}\bar{c}^{b}c^{c}c^{d}\right.\nonumber\\
&&\phantom{+\alpha\int
d^{4}\!x\,}\left.-\frac{g^{2}}{4}f^{abc}f^{adi}\bar{c}^{b}\bar{c}^{c}c^{d}c^{i}
-\frac{g^{2}}{8}f^{abc}f^{ade}\bar{c}^{b}\bar{c}^{c}c^{d}c^{e}\right)
\;,\label{MAG}
\end{eqnarray}
where $b^A \equiv (b^a,b^i)$ are the Nakanishi-Lautrup fields, and
$c^A  = (c^a,c^i)$, ${\bar c}^A \equiv ({\bar c}^a, {\bar c}^i)$
are the ghost and antighost fields. The gauge parameter $\alpha$
is introduced in (\ref{MAG}) for renormalizability purposes. As a
consequence of the nonlinearity of condition (\ref{MAGgf}), the
quartic interaction ghost terms in expression (\ref{MAG}) is
required in order to obtain a stable action
\cite{Dudal:2004rx,Min:1985bx,Fazio:2001rm}. After the removal of
the ultraviolet divergences, the limit $\alpha \rightarrow 0$ has
to be considered in order to achieve (\ref{MAGgf}). \\\\In
non-Abelian gauge theories one has to face the existence of the
Gribov ambiguities \cite{Gribov:1977wm}, which deeply affect the
infrared region. In the MAG, it is known that condition
(\ref{MAGgf}) does not uniquely fix the gauge
\cite{Bruckmann:2000xd}, so that a suitable restriction of the
domain of integration in the Feynman path integral has to be
implemented in order to avoid the counting of equivalent field
configurations. The renormalization of the operator
(\ref{trmassop}) has already been investigated in the Landau gauge
\cite{Capri:2007ix} when the restriction of the domain of
integration to the so called Gribov region was taken into account.
In \cite{Capri:2007ix}, it was explicitly shown that the
introduction of the Zwanziger horizon function
\cite{Zwanziger:1989mf,Zwanziger:1992qr}, which implements the
restriction to the Gribov region, does not spoil the
renormalizability of (\ref{trmassop}). The same feature occurs
here in the MAG. However, for simplicity, we have decided of not
including in the main text the lengthy and technical analysis of
the mass operator (\ref{trmassop}) in the presence of the horizon
function of the MAG, leaving the inclusion of the horizon term to
Appendix {\bf A}, where the interested reader may find a detailed
discussion.

\subsection{Localizing the mass operator}
In \cite{Capri:2005dy}, it has been shown that the nonlocal mass
term (\ref{trmassop}) may be written in a local form by means of
the introduction of a pair of complex bosonic antisymmetric tensor
fields in the adjoint representation,
$(B^A_{\mu\nu},{\bar{B}}^A_{\mu\nu})$, and a pair of anticommuting
antisymmetric complex tensor fields,
$(G^A_{\mu\nu},{\bar{G}}^A_{\mu\nu})$. Following
\cite{Capri:2005dy}, the nonlocal operator ${\cal O}$,
eq.(\ref{trmassop}), is coupled to the Yang-Mills action by
introducing the gauge invariant mass term
\begin{equation}
 S_{\cal O} = -{m^2 \over 4} \int d^4\!x\, F_{\mu\nu}^A
 {\left[{(D^2)}^{-1}\right]}^{AB} F_{\mu\nu}^B
 \;.  \label{Smassop}
 \end{equation}
Furthermore, it is easily checked that expression (\ref{Smassop})
can be rewritten in local form as
\begin{equation}
 e^{-S_{\cal O}} =
 \int D{\bar B} DB D{\bar G} D G  \exp \left[ -(S_{BG} + S_m) \right]
 \label{idy}
 \;,
\end{equation}
with
\begin{eqnarray}
 S_{BG}&=& {1 \over 4} \int{d^{4}\!x} \Bigl( \bar{B}^{A}_{\mu\nu}
 {\cal O}^{AB} B^B_{\mu\nu}
 - \bar{G}^{A}_{\mu\nu} {\cal O}^{AB} G^B_{\mu\nu}
 \Bigr)
 %S_{BG}&=&\int{d^{4}\!x}\,
%\biggl\{\bar{B}^{a}_{I}\mathcal{O}^{ab}B^{b}_{I}
%-\bar{G}^{a}_{I}\mathcal{O}^{ab}G^{b}_{I}
%+\bar{B}^{a}_{I}\mathcal{O}^{ai}B^{i}_{I}
%-\bar{G}^{a}_{I}\mathcal{O}^{ai}G^{i}_{I}
% \biggr.\nonumber\\&&\biggl.
%+\bar{B}^{i}_{I}\mathcal{O}^{ia}B^{a}_{I}
%-\bar{G}^{i}_{I}\mathcal{O}^{ia}G^{a}_{I}
%+\bar{B}^{i}_{I}\mathcal{O}^{ij}B^{j}_{I}
%-\bar{G}^{i}_{I}\mathcal{O}^{ij}G^{j}_{I}
% \biggr\}
 \;,
 \\
 S_m &=& {im \over 4} \int d^4\!x\, (B^A_{\mu\nu} - {\bar B}^A_{\mu\nu})
 F^A_{\mu\nu}
 \label{Sm}
\;,
\end{eqnarray}
and
\begin{equation}
\mathcal{O}^{AB}\equiv
%\left(\begin{matrix}
%\mathcal{O}^{ab}&\mathcal{O}^{aj}\cr
%\mathcal{O}^{ib}&\mathcal{O}^{ij}
%\end{matrix}\right)=
D^{AC}_{\mu}D^{CB}_{\mu}\label{local_op} \;.
\end{equation}
The identity \eqref{idy} allows us to localize the expression
$S_{\cal O}$, \eqref{Smassop}, when added to the YM action
$S_{\mathrm{YM}}$, eq.\eqref{YM}. Thus, for the local gauge-fixed
action $S_{\mathrm{phys}}$ in the MAG we write
\begin{equation}
 S_{\mathrm{phys}} = S_{\mathrm{YM}} + S_{\mathrm{MAG}} + S_{BG} + S_m
 \;.
 \label{Sphys}
\end{equation}
Evidently
\begin{equation}
 \int D{\bar B} DB D{\bar G} D G  \exp \left[ -S_{\mathrm{phys}} \right]
 = e^{ -( S_{\mathrm{YM}}+S_{\mathrm{MAG}}+S_{\cal O} )}
\,.
\end{equation}

\subsection{BRST invariance}
In order to establish the BRST invariance of the local action
$S_{\mathrm{phys}}$, we shall employ here the same procedure of
\cite{Zwanziger:1992qr,Capri:2007ck,Capri:2005dy,Capri:2006cz},
and we shall embed the action $S_{\mathrm{phys}}$ into a more
general one. Following
\cite{Zwanziger:1992qr,Capri:2007ck,Capri:2005dy,Capri:2006cz}, we
introduce the system of external sources
\begin{equation}
 (U_{\sigma\rho\mu\nu},\bar{U}_{\sigma\rho\mu\nu},
V_{\sigma\rho\mu\nu},\bar{V}_{\sigma\rho\mu\nu}) \label{extsources}
\end{equation}
and replace the term $S_m$ in (\ref{Sphys}) by
\begin{equation}
S_{UV} = {1 \over 4} \int d^4x
F^{A}_{\mu\nu}(\bar{U}_{\lambda\rho\mu\nu}G^{A}_{\lambda\rho}
+V_{\lambda\rho\mu\nu}\bar{B}^{A}_{\lambda\rho}
 -\bar{V}_{\lambda\rho\mu\nu}B^{A}_{\lambda\rho}
+U_{\lambda\rho\mu\nu}\bar{G}^{A}_{\lambda\rho}) \label{SUV} \;.
\end{equation}
Notice that expression (\ref{Sm}) is recovered from (\ref{SUV}) in
the physical limit for the sources, namely
\begin{eqnarray}
\bar{V}_{\sigma\rho\mu\nu}\Bigl|_{\mathrm{phys}}&=&
V_{\sigma\rho\mu\nu}\Bigl|_{\mathrm{phys}}
=-\frac{im}{2}(\d_{\sigma\mu}\d_{\rho\nu}
-\d_{\sigma\nu}\d_{\rho\nu})\;,\nonumber\\
U_{\sigma\rho\mu\nu}\Bigl|_{\mathrm{phys}}&=&
\bar{U}_{\sigma\rho\mu\nu}\Bigl|_{\mathrm{phys}}=0\;,
\label{phys_limit}
\end{eqnarray}
i.e.
\begin{equation}
S_{UV}\Bigl|_{\mathrm{phys}} = S_m \label{new} \;.
\end{equation}
Thus, action (\ref{Sphys}) is replaced by
\begin{equation}
 S_{\mathrm{inv}} = S_{\mathrm{YM}} + S_{\mathrm{MAG}} + S_{BG} +
 S_{UV}
 \label{Sinv}
 \;,
\end{equation}
which defines a more general theory and has $S_{\mathrm{phys}}$ as
a particular case. \\\\As an important bonus, we observe that
$S_{\mathrm{inv}}$ is invariant under the following global $U(6)$
symmetry
\begin{equation}
\mathcal{Q}_{\mu\nu\lambda\rho}(S_{\mathrm{inv}})=0\;,\label{u(6)inc}
\end{equation}
with
\begin{eqnarray}
\mathcal{Q}_{\mu\nu\lambda\rho}&\!\!\!\equiv\!\!\!&\Int\left(B^{A}_{\mu\nu}\frac{\d}{\d
B^{A}_{\lambda\rho}}
 -\bar{B}^{A}_{\lambda\rho}\frac{\d}{\d\bar{B}^{A}_{\mu\nu}}
+G^{A}_{\mu\nu}\frac{\d}{\d G^{A}_{\lambda\rho}}
-\bar{G}^{A}_{\lambda\rho}\frac{\d}{\d\bar{G}^{A}_{\mu\nu}}\right.\nonumber\\
&&\left. +U_{\mu\nu\alpha\beta}\frac{\d}{\d
U_{\lambda\rho\alpha\beta}}
-\bar{U}_{\lambda\rho\alpha\beta}\frac{\d}{\d\bar{U}_{\mu\nu\alpha\beta}}+V_{\mu\nu\alpha\beta}\frac{\d}{\d
 V_{\lambda\rho\alpha\beta}}
-\bar{V}_{\lambda\rho\alpha\beta}\frac{\d}{\d\bar{V}_{\mu\nu\alpha\beta}}
\right)\;.\label{Q_opinc}
\end{eqnarray}
The symmetry (\ref{u(6)inc}) is naturally related to the mass
operator and allows us to use a multi-index notation,
$(\mu,\nu)\rightarrow I$, $ I = 1,\dots,6 $, so that
\begin{eqnarray}
(B_{I}^A,\bar{B}_{I}^A,G_{I}^A,\bar{G}_{I}^A)&=&
\frac{1}{2}(B_{\mu\nu}^A,\bar{B}_{\mu\nu}^A,
G_{\mu\nu}^A,\bar{G}_{\mu\nu}^A)\;,
\\
(U_{I\mu\nu},\bar{U}_{I\mu\nu},V_{I\mu\nu},\bar{V}_{I\mu\nu})&=&
\frac{1}{2}(U_{\sigma\rho\mu\nu},\bar{U}_{\sigma\rho\mu\nu},
V_{\sigma\rho\mu\nu},\bar{V}_{\sigma\rho\mu\nu})\;.
\end{eqnarray}
The use of the multi-index $I$ will turn out to be very useful
when looking for combinations of possible counterterms respecting
(\ref{u(6)inc}). With this notation, we may rewrite the action
(\ref{Sinv}) as
\begin{eqnarray}
 S_{\mathrm{inv}} &=& S_{\mathrm{YM}} + S_{\mathrm{MAG}} +
   \int{d^{4}\!x} ( \bar{B}^{A}_{I} {\cal O}^{AB} B^B_{I}
 - \bar{G}^{A}_{I} {\cal O}^{AB} G^B_{I}
   \nonumber\\
&+& F^{A}_{\mu\nu}(\bar{U}_{I\mu\nu}G^{A}_{I}
+V_{I\mu\nu}\bar{B}^{A}_{I} -\bar{V}_{I\mu\nu}B^{A}_{I}
+U_{I\mu\nu}\bar{G}^{A}_{I}) ) \label{SINV} \;.
\end{eqnarray}
The introduction of the system of sources (\ref{extsources})
enables us to establish that the action (\ref{SINV}) possesses a
BRST invariance. In fact, it can be checked  by direct inspection
that the following nilpotent transformations
\begin{eqnarray}
sA^{a}_{\mu}&=&-(D^{ab}_{\mu}c^{b} +gf^{abc}A^{b}_{\mu}c^{c}
+gf^{abi}A^{b}_{\mu}c^{i})\;,\nonumber\\
sA^{i}_{\mu}&=&-(\partial_{\mu}c^{i}+gf^{abi}A^{a}_{\mu}c^{b})\;,\nonumber\\
sc^{a}&=&gf^{abi}c^{b}c^{i}+\frac{g}{2}f^{abc}c^{b}c^{c}\;,\nonumber\\
sc^{i}&=&\frac{g}{2}f^{abi}c^{a}c^{b}\;,\nonumber\\
s\bar{c}^{A}&=&b^{A\phantom{i}}\;,\qquad
sb^{A\phantom{i}}=\,0\;,\nonumber\\
sB^{A}_{I}&=&G^{A}_{I}+gf^{ABC}c^{B}B^{C}_{I}\;,\nonumber\\
s\bar{B}^{A}_{I}&=&gf^{ABC}c^{B}\bar{B}^{C}_{I}\;,\nonumber\\
sG^{A}_{I}&=&gf^{ABC}c^{B}G^{C}_{I}\;,\nonumber\\
s\bar{G}^{A}_{I}&=&\bar{B}^{A}_{I}+gf^{ABC}c^{B}\bar{G}^{C}_{I}\;,\nonumber\\
sV_{I\mu\nu}&=&U_{I\mu\nu}\;,\qquad sU_{I\mu\nu}=0\;,\nonumber\\
s\bar{U}_{I\mu\nu}&=&\bar{V}_{I\mu\nu}\;,\qquad\hspace{2.5pt}
s\bar{V}_{I\mu\nu}=0\;.\label{brst}
\end{eqnarray}
leave the action (\ref{SINV}) invariant.  In particular, we can
 rewrite
\begin{equation}
 S_{\mathrm{MAG}} = s \Int
\left[ {\bar c}^a \left(D^{ab}_\mu + {\alpha\over2}b^a\right)
-{\alpha\over2}gf^{abi}{\bar c}^a{\bar c}^b c^i
-{\alpha\over4}gf^{abc}c^a{\bar c}^b{\bar c}^c +{\bar
c}^i\partial_\mu A^i_\mu \right] \;.
\end{equation}

%%%%%%%%%%%%%%%%%%%%%%%%%%
\section{Identification of the complete starting action and its
 symmetries}
%%%%%%%%%%%%%%%%%%%%%%%%%%

In order to prove the renormalizability of the action
$S_{\mathrm{inv}}$, eq.(\ref{SINV}), we shall make use of the
algebraic renormalization technique \cite{Piguet:1995er}. To that
purpose, we shall introduce a set of BRST sources and we shall
establish the Ward identities fulfilled by the theory. The
knowledge of these Ward identities will play a central role in the
determination of the most general allowed invariant counterterm.

\subsection{Starting action}
As is well known, due to the nonlinear character of the BRST
transformations, eqs.\eqref{brst}, one has to introduce a set of
external sources, $(X^A_I,{\bar X}^A_I,Y^A_I,{\bar
 Y}^A_I,\Omega^A_\mu,L^A) $, coupled to them, namely
 \begin{eqnarray}
S_{\mathrm{ext}}&\!\!\!=\!\!\!&\Int\,\biggl[-\Omega^{a}_{\mu}(D^{ab}_{\mu}c^{b}
+gf^{abc}A^{b}_{\mu}c^{c}
+gf^{abi}A^{b}_{\mu}c^{i})-\Omega^{i}_{\mu}(\partial_{\mu}c^{i}
+gf^{abi}A^{a}_{\mu}c^{b}) +L^{a}\biggl(gf^{abi}c^{b}c^{i}
+\frac{g}{2}f^{abc}c^{b}c^{c}\biggr) \nonumber\\&&
+\frac{g}{2}f^{abi}L^{i}c^{a}c^{b}
+\bar{Y}^{A}_{I}gf^{ABC}c^{B}B^{C}_{I}
+Y^{A}_{I}gf^{ABC}c^{B}\bar{B}^{C}_{I}
+\bar{X}^{A}_{I}gf^{ABC}c^{B}G^{C}_{I}
+X^{A}_{I}gf^{ABC}c^{B}\bar{G}^{C}_{I}
\biggr]\;,\label{external}
\end{eqnarray}
Further, we introduce the following two terms, $S_{\lambda}$ and
$S_{\mathrm{sources}}$ to be added to the action
$S_{\mathrm{inv}}$:
\begin{eqnarray}
S_{\lambda}&\!\!\!=\!\!\!&\Int\,\biggl\{\lambda_{1}(\bar{B}^{A}_{I}B^{A}_{I}
-\bar{G}^{A}_{I}G^{A}_{I}) (\bar{V}_{J\mu\nu}V_{J\mu\nu}
-\bar{U}_{J\mu\nu}U_{J\mu\nu})
\nonumber\\
&&+\lambda_{2}\biggl(\bar{B}^{A}_{I}G^{A}_{J}
V_{I\mu\nu}\bar{U}_{J\mu\nu}
+\bar{G}^{A}_{I}G^{A}_{J}U_{I\mu\nu}\bar{U}_{J\mu\nu}
+\bar{B}^{A}_{I}B^{A}_{J}V_{I\mu\nu}\bar{V}_{J\mu\nu}
-\bar{G}^{A}_{I}B^{A}_{J}U_{I\mu\nu}\bar{V}_{J\mu\nu}
-G^{A}_{I}B^{A}_{J}\bar{U}_{I\mu\nu}\bar{V}_{J\mu\nu}\nonumber\\
&&+\bar{G}^{A}_{I}\bar{B}^{A}_{J}U_{I\mu\nu}V_{J\mu\nu}
-\frac{1}{2}B^{A}_{I}B^{A}_{J}\bar{V}_{I\mu\nu}\bar{V}_{J\mu\nu}
+\frac{1}{2}G^{A}_{I}G^{A}_{J}\bar{U}_{I\mu\nu}\bar{U}_{J\mu\nu}
-\frac{1}{2}\bar{B}^{A}_{I}\bar{B}^{A}_{J}V_{I\mu\nu}V_{J\mu\nu}
+\frac{1}{2}\bar{G}^{A}_{I}\bar{G}^{A}_{J}U_{I\mu\nu}U_{J\mu\nu}\biggr)\nonumber\\
&&+\frac{\lambda^{ABCD}}{16}(\bar{B}^{A}_{I}B^{B}_{I}-\bar{G}^{A}_{I}G^{B}_{I})
(\bar{B}^{C}_{J}B^{D}_{J}-\bar{G}^{C}_{J}G^{D}_{J})
\biggr\}\;,\label{lambda_term}\\\cr
S_{\mathrm{sources}}&\!\!\!=\!\!\!&\Int\,\Bigl[\chi_{1}(\bar{V}_{I\mu\nu}\p^{2}V_{I\mu\nu}
-\bar{U}_{I\mu\nu}\p^{2}U_{I\mu\nu})
+\chi_{2}(\bar{V}_{I\mu\nu}\p_{\mu}\p_{\s}V_{I\nu\s}
-\bar{U}_{I\mu\nu}\p_{\mu}\p_{\s}U_{I\nu\s})\nonumber\\
&&-\zeta(\bar{U}_{I\mu\nu}U_{I\mu\nu}\bar{U}_{J\s\rho}U_{J\s\rho}
+\bar{V}_{I\mu\nu}V_{I\mu\nu}\bar{V}_{J\s\rho}V_{J\s\rho}
-2\bar{U}_{I\mu\nu}U_{I\mu\nu}\bar{V}_{J\s\rho}V_{J\s\rho})\Bigr]\;.\label{sources}
\end{eqnarray}
The first term, $S_{\lambda}$, already discussed in detail in
\cite{Capri:2005dy,Capri:2006ne}, contains interactions between
the auxiliary fields and the sources $(U_{I\mu\nu},
\bar{U}_{I\mu\nu}, V_{I\mu\nu}\bar{V}_{I\mu\nu})$, and is needed
for the stability of the action.  The second term,
$S_{\mathrm{sources}}$, depends only from $(U_{I\mu\nu},
\bar{U}_{I\mu\nu}, V_{I\mu\nu}\bar{V}_{I\mu\nu})$ and is allowed
by power counting. The parameters $\lambda_{1}$, $\lambda_{2}$,
$\chi_{1}$, $\chi_{2}$ and $\zeta$ are free, while the 4-rank
invariant tensor $\lambda^{ABCD}$ enjoys the properties
\cite{Capri:2005dy}
\begin{equation}
f^{MAN}\lambda^{MBCD}+ f^{MBN}\lambda^{AMCD} +f^{MCN}\lambda^{ABMD}
+f^{MDN}\lambda^{ABCM}=0\;,
\end{equation}
and
\begin{equation}
\lambda^{ABCD}=\lambda^{CDAB}=\lambda^{BACD}\;.
\end{equation}
Collecting all terms, we finally write the starting action
$\Sigma$ as
\begin{equation}
\Sigma=S_{\mathrm{inv}}+ S_{\lambda}+S_{\mathrm{ext}}
+S_{\mathrm{sources}}\;. \label{Sigma}
\end{equation}
The stability of $\Sigma$ under quantum corrections will be
investigated in the next section. \\\\Here, for the sake of
clarity, we recall the range of variations of all the indices
introduced so far;
\begin{eqnarray}
A,B,C,D,\ldots&\in&\{1,\ldots,N^{2}-1\}\;,\nonumber\\
a,b,c,d,\ldots&\in&\{1,\ldots,N(N-1)\}\;,\nonumber\\
i,j,k,l,\ldots&\in&\{1,\ldots,N-1\}\;,\nonumber\\
I,J,K,L\ldots&\in&\{1,\ldots,\half D(D-1)=6\}\;,\nonumber\\
\mu,\nu,\sigma,\rho,\ldots&\in&\{1,\ldots,D=4\}\;.\label{set_indices}
\end{eqnarray}

\begin{table}[t] \centering
{\small\begin{tabular}{lcccccccccccccccccc} \hline\hline
&$A$&$b$&$\bar c$&$c$&$B$&$\bar{B}$&$G$&$\bar{G}$&$U$&$\bar{U}$&
$V$&$\bar{V}$&$Y$&$\bar{Y}$&$X$&$\bar{X}$&$\Omega$&$L\!\phantom{\Bigl|}$\\
\hline
dim&$1$&$2$&$2$&$0$&$1$&$1$&$1$&$1$&$1$&$1$&$1$&$1$&$3$&$3$&$3$&$3$&$3$&$4$\\
gh.
number&$0$&$0$&$-1$&$1$&$0$&$0$&$1$&$-1$&$1$&$-1$&$0$&$0$&$-1$&$-1$&$-2$&
$0$&$-1$&$-2$\\
$\mathcal{Q}_{6}$-charge&$0$&$0$&$0$&$0$&$1$&$-1$&$1$&$-1$&$1$&$-1$&$1$&$-1$
&$1$&$-1$&$1$&$-1$&$0$&$0$\\
\hline\hline
\end{tabular}}
\caption{Quantum numbers of the fields and sources} \label{table1}
\end{table}
\noindent In table (\ref{table1}) we display the dimensions and
the ghost number of the complete set of fields and sources of the
theory, as well as the corresponding ${\cal Q}_6$-charge, which is
defined as the trace of the of the operator (\ref{Q_opinc}).
\\\\The starting action (\ref{Sigma}) is left invariant by the
action of the nilpotent BRST operator $s$ given by (\ref{brst})
and by
\begin{eqnarray}
s\Omega^{A}_{\mu}&=&0\;,\qquad\hspace{6.5pt} sL^{A}=0\;,\nonumber\\
sY^{A}_{I}&=&X^{A}_{I}\;,\qquad\hspace{6.5pt}
 sX^{A}_{I}=0\;,\nonumber\\
s\bar{X}^{A}_{I}&=&-\bar{Y}^{A}_{I}\;,\qquad
s\bar{Y}^{A}_{I}=0\;,\label{brst_sources}
\end{eqnarray}
{\it i.e.}
\begin{equation}
s \Sigma = 0 \ . \label{new2}
\end{equation}

%%%%%%%%%%%%%%%%%%%%%%%%%
\subsection{Ward identities}
%%%%%%%%%%%%%%%%%%%%%%%%%
In order to constrain the possible counterterms which can be added
to the local starting action $\Sigma$, eq.(\ref{Sigma}), let us
proceed by establishing the set of Ward identities fulfilled by
the starting classical action. In fact, it turns out that $\Sigma$
obeys the following set of Ward identities:
\begin{itemize}
\item{The diagonal gauge fixing identity:
\begin{equation}
\frac{\d\S}{\d b^{i}}=\partial_{\mu}A^{i}_{\mu}\;.\label{diag_gf}
\end{equation}}

\item{The diagonal anti-ghost equation:
\begin{equation}
\frac{\d\S}{\d\bar{c}^{i}}+\p_{\mu}\frac{\d\S}{\d\Omega^{i}_{\mu}}=0\;.
\label{diag_anti_gh}
\end{equation}}

\item{The diagonal ghost equation:
\begin{equation}
\mathcal{G}^{i}(\S)=\Delta^{i}_{\mathrm{class}}\;,\label{diag_gh}
\end{equation}
where
\begin{equation}
\mathcal{G}^{i}= \frac{\d}{\d c^{i}}+gf^{abi}\bar{c}^{a}
\frac{\d}{\d b^{b}}\;,\label{gh_op}
\end{equation}
and
\begin{eqnarray}
\Delta^{i}_{\mathrm{class}}&=&-\p^{2}\bci -\p_{\mu}\Omega^{i}
+gf^{abi}\Omega^{a}_{\mu}A^{b}_{\mu} -gf^{abi}L^{a}c^{b}
+gf^{abi}\bar{Y}^{a}_{I}B^{b}_{I}\nonumber\\
&&+gf^{abi}Y^{a}_{I}\bar{B}^{b}_{I}
-gf^{abi}\bar{X}^{a}_{I}G^{b}_{I}
-gf^{abi}X^{a}_{I}\bar{G}^{b}_{I}\;.\label{class_breaking}
\end{eqnarray}}
Notice that the term $\Delta^{i}_{\mathrm{class}}$, being linear
in the quantum fields, represents a classical breaking not
affected by quantum corrections \cite{Piguet:1995er}.

\item{The Slavnov-Taylor identity
\begin{equation}
\mathcal{S}(\S)=0\;,\label{ST}
\end{equation}
where
\begin{eqnarray}
\mathcal{S}(\S)&\equiv&\Int\,\left[
\frac{\d\S}{\d\Omega^{A}_{\mu}}\frac{\d\S}{\d A^{A}_{\mu}}
+\frac{\d\S}{\d L^{A}}\frac{\d\S}{\d c^A} +b^{A}\frac{\d\S}{\d{\bar
 c}^A}
+\left(\frac{\d\S}{\d\bar{Y}^{A}_{I}}+G^{A}_{I}\right)\frac{\d\S}{\d
B^{A}_{I}} +\frac{\d\S}{\d Y^{A}_{I}}\frac{\d\S}{\d\bar{B}^{A}_{I}}
+\frac{\d\S}{\d\bar{X}^{A}_{I}}\frac{\d\S}{\d G^{A}_{I}}
\right.\nonumber\\
&& +\left.\left(\frac{\d\S}{\d
X^{A}_{I}}+\bar{B}^{A}_{I}\right)\frac{\d\S}{\d\bar{G}^{A}_{I}}
+\bar{V}_{I\mu\nu}\frac{\d\S}{\d\bar{U}_{I\mu\nu}}
+U_{I\mu\nu}\frac{\d\S}{\d V_{I\mu\nu}}
-\bar{Y}^{A}_{I}\frac{\d\S}{\d\bar{X}^{A}_{I}}
+{X}^{A}_{I}\frac{\d\S}{\d{Y}^{A}_{I}} \right]\;.\label{S}
\end{eqnarray}
The Slavnov-Taylor identity \eqref{ST} gives rise to the
corresponding nilpotent linearized operator
\begin{eqnarray}
\mathcal{S}_{\S}&\!\!\!\equiv\!\!\!&\Int\left[\frac{\d\S}{\d\Omega^{A}_{\mu}}\frac{\d}{\d
A^{A}_{\mu}} +\frac{\d\S}{\d{A}^{A}_{\mu}}\frac{\d}{\d
\Omega^{A}_{\mu}} + \frac{\d\S}{\d L^{A}}\frac{\d}{\d c^A}
+\frac{\d\S}{\d
 c^A}\frac{\d}{\d{L}^{A}}+b^{A}\frac{\d}{\d\bar{c}^A}\right.\nonumber\\
&&+\left(\frac{\d\S}{\d\bar{Y}^{A}_{I}}+G^{A}_{I}\right)\frac{\d}{\d
B^{A}_{I}}+\frac{\d\S}{\d B^{A}_{I}}\frac{\d}{\d\bar{Y}^{A}_{I}}
+\frac{\d\S}{\d Y^{A}_{I}}\frac{\d}{\d\bar{B}^{A}_{I}}
+\left(\frac{\d\S}{\d\bar{B}^{A}_{I}}+X^{A}_{I}\right)\frac{\d}{\d
Y^{A}_{I}}\nonumber\\
&&+\frac{\d\S}{\d\bar{X}^{A}_{I}}\frac{\d}{\d G^{A}_{I}}
+\frac{\d\S}{\d\bar{X}^{A}_{I}}\frac{\d}{\d G^{A}_{I}}
+\left(\frac{\d\S}{\d{G}^{A}_{I}}-\bar{Y}^{A}_{I}\right)\frac{\d}{\d
\bar{X}^{A}_{I}}+\left(\frac{\d\S}{\d{X}^{A}_{I}}+\bar{B}^{A}_{I}\right)\frac{\d}{\d
\bar{G}^{A}_{I}}\nonumber\\
&&\left.+\frac{\d\S}{\d\bar{G}^{A}_{I}}\frac{\d}{\d X^{A}_{I}}
+\bar{V}_{I\mu\nu}\frac{\d}{\d\bar{U}_{I\mu\nu}}
+U_{I\mu\nu}\frac{\d}{\d V_{I\mu\nu}}\right]\;,\label{linearized}
\end{eqnarray}}
\begin{equation}
\mathcal{S}_{\S} \mathcal{S}_{\S} =0 \;. \label{new3}
\end{equation}
\item{The diagonal $U(1)^{N-1}$ Ward identity:
\begin{equation}
\mathcal{W}^{i}(\S)=-\p^{2}b^{i}\;,\label{u(1)}
\end{equation}
with
\begin{eqnarray}
\mathcal{W}^{i}&\!\!\!\equiv\!\!\!&\p_{\mu}\frac{\d}{\d
A^{i}_{\mu}}+gf^{abi}\left(A^{a}_{\mu}\frac{\d}{\d{A}^{b}_{\mu}}
+b^{a}\frac{\d}{\d b^{b}} +c^{a}\frac{\d}{\d c^{b}}
+\bar{c}^{a}\frac{\d}{\d \bar{c}^{b}} +B^{a}_{I}\frac{\d}{\d
B^{b}_{I}} +\bar{B}^{a}_{I}\frac{\d}{\d
\bar{B}^{b}_{I}}\right.\nonumber\\
&&\left.\!+G^{a}_{I}\frac{\d}{\d G^{b}_{I}}
+\bar{G}^{a}_{I}\frac{\d}{\d \bar{G}^{b}_{I}}
+\Omega^{a}_{\mu}\frac{\d}{\d \Omega^{b}_{\mu}} +L^{a}\frac{\d}{\d
L^{b}} +Y^{a}_{I}\frac{\d}{\d Y^{b}_{I}}
+\bar{Y}^{a}_{I}\frac{\d}{\d \bar{Y}^{b}_{I}} +X^{a}_{I}\frac{\d}{\d
X^{b}_{I}} +\bar{X}^{a}_{I}\frac{\d}{\d \bar{X}^{b}_{I}}\right)\;.
\label{u(1)_op}
\end{eqnarray}}

\item{The off-diagonal $SL(2,\mathbb{R})$ identity:
\begin{equation}
\mathcal{D}(\Sigma)=0\;,\label{sl2r}
\end{equation}
where
\begin{equation}
\mathcal{D}(\Sigma)=\;\Int\,\left(c^{a}\frac{\d\S}{\d\bar{c}^{a}}
+\frac{\d\S}{\d L^{a}}\frac{\d\S}{\d b^{a}}\right)\;.
\end{equation}
As the Slavnov-Taylor identity, the equation (\ref{sl2r}) also defines
 a linear operator:
\begin{equation}
\mathcal{D}_{\S}\equiv\;\Int\,\left(c^{a}\frac{\d}{\d\bar{c}^{a}}
+\frac{\d\S}{\d L^{a}}\frac{\d}{\d b^{a}} +\frac{\d\S}{\d
b^{a}}\frac{\d}{\d L^{a}}\right)\;.\label{sl2r_op}
\end{equation}}

\item{The global $U(6)$ invariance related to the nonlocal mass
operator:
\begin{equation}
\mathcal{Q}_{IJ}(\Sigma)=0\;,\label{u(6)}
\end{equation}
where
\begin{eqnarray}
\mathcal{Q}_{IJ}&\!\!\!\equiv\!\!\!&\Int\left(B^{A}_{I}\frac{\d}{\d
B^{A}_{J}} -\bar{B}^{A}_{J}\frac{\d}{\d\bar{B}^{A}_{I}}
+G^{A}_{I}\frac{\d}{\d G^{A}_{J}}
-\bar{G}^{A}_{J}\frac{\d}{\d\bar{G}^{A}_{I}}
+U_{I\mu\nu}\frac{\d}{\d U_{J\mu\nu}}
-\bar{U}_{J\mu\nu}\frac{\d}{\d\bar{U}_{I\mu\nu}}\right.\nonumber\\
&&\left.+V_{I\mu\nu}\frac{\d}{\d V_{J\mu\nu}}
-\bar{V}_{J\mu\nu}\frac{\d}{\d\bar{V}_{I\mu\nu}}
+Y^{A}_{I}\frac{\d}{\d Y^{A}_{J}}
-\bar{Y}^{A}_{J}\frac{\d}{\d\bar{Y}^{A}_{I}} +X^{A}_{I}\frac{\d}{\d
X^{A}_{J}}
-\bar{X}^{A}_{J}\frac{\d}{\d\bar{X}^{A}_{I}}\right)\;.\label{Q_op}
\end{eqnarray}
The trace of \eqref{Q_op}} defines the
$\mathcal{Q}_{6}$\emph{-charge},
$\mathcal{Q}_{6}=\mathcal{Q}_{II}$, already displayed in table
(\ref{table1}).

\item{The exact rigid symmetries:
\begin{equation}
\mathcal{R}^{(\mathcal{N})}_{IJ}(\Sigma)=0\;,\label{rigid}
\end{equation}
being $\mathcal{N}=1,2,3,4$ and
\begin{eqnarray}
\mathcal{R}^{(1)}_{IJ}&\equiv&\Int\left(B^{A}_{I}\frac{\d}{\d{G}^{A}_{J}}
-\bar{G}^{A}_{J}\frac{\d}{\d\bar{B}^{A}_{I}}
+V_{I\mu\nu}\frac{\d}{\d{U}_{J\mu\nu}}
-\bar{U}_{J\mu\nu}\frac{\d}{\d\bar{V}_{I\mu\nu}}
+Y^{A}_{I}\frac{\d}{\d{X}^{A}_{J}}
+\bar{X}^{A}_{J}\frac{\d}{\d\bar{Y}^{A}_{I}}\right)\;,\nonumber\\
\mathcal{R}^{(2)}_{IJ}&\equiv&\Int\left(\bar{B}^{A}_{I}\frac{\d}{\d\bar{G}^{A}_{J}}
+{G}^{A}_{J}\frac{\d}{\d{B}^{A}_{I}}
+\bar{V}_{I\mu\nu}\frac{\d}{\d\bar{U}_{J\mu\nu}}
+{U}_{J\mu\nu}\frac{\d}{\d{V}_{I\mu\nu}}
-\bar{Y}^{A}_{I}\frac{\d}{\d\bar{X}^{A}_{J}}
+{X}^{A}_{J}\frac{\d}{\d{Y}^{A}_{I}}\right)\;,\nonumber\\
\mathcal{R}^{(3)}_{IJ}&\equiv&\Int\left(\bar{B}^{A}_{I}\frac{\d}{\d{G}^{A}_{J}}
-\bar{G}^{A}_{J}\frac{\d}{\d{B}^{A}_{I}}
-\bar{V}_{I\mu\nu}\frac{\d}{\d{U}_{J\mu\nu}}
+\bar{U}_{J\mu\nu}\frac{\d}{\d{V}_{I\mu\nu}}
+\bar{Y}^{A}_{I}\frac{\d}{\d{X}^{A}_{J}}
+\bar{X}^{A}_{J}\frac{\d}{\d{Y}^{A}_{I}}\right)\;,\nonumber\\
\mathcal{R}^{(4)}_{IJ}&\equiv&\Int\left(B^{A}_{I}\frac{\d}{\d\bar{G}^{A}_{J}}
+{G}^{A}_{J}\frac{\d}{\d\bar{B}^{A}_{I}}
-V_{I\mu\nu}\frac{\d}{\d\bar{U}_{J\mu\nu}}
-{U}_{J\mu\nu}\frac{\d}{\d\bar{V}_{I\mu\nu}}
-Y^{A}_{I}\frac{\d}{\d\bar{X}^{A}_{J}}
+{X}^{A}_{J}\frac{\d}{\d\bar{Y}^{A}_{I}}\right)\;.\label{rigit_ops}
\end{eqnarray}}
\end{itemize}
%%%%%%%%%%%%%%%%%%%%%%%%%%%%%%%%%%%%%%%%%%%%%%%%%%%%%%%%%%%
%\section{Useful commutation and anti-commutation relations}
%%%%%%%%%%%%%%%%%%%%%%%%%%%%%%%%%%%%%%%%%%%%%%%%%%%%%%%%%%%
For calculation purposes, we display the following useful
(anti-)commutation relations between the linearized operator
$\mathcal{S}_{\Sigma}$, given in \eqref{linearized}, and the
operators \eqref{gh_op}, \eqref{sl2r_op}, and \eqref{rigit_ops},
namely
\begin{eqnarray}
\Bigl[\mathcal{D}_{\S},\mathcal{S}_{\Sigma}\Bigr]=0\;,&&
\Bigl\{\mathcal{G}^{i},\mathcal{S}_{\Sigma}\Bigr\}=\mathcal{W}^{i}\;,\\
\Bigl\{\mathcal{R}^{(1)}_{IJ},\mathcal{S}_{\Sigma}\Bigr\}=\mathcal{Q}_{IJ}\;,&&
\Bigl\{\mathcal{R}^{(2)}_{IJ},\mathcal{S}_{\Sigma}\Bigr\}=0\;,
\end{eqnarray}
\begin{eqnarray}
\Bigl\{\mathcal{R}^{(3)}_{IJ},\mathcal{S}_{\Sigma}\Bigr\}&=& \int
d^{4}\!x\,\left(\delta_{IK}\delta_{JL}-\delta_{IL}\delta_{JK}\right)
\left(\bar B^{A}_{K}\frac{\delta}{\delta B^{A}_{L}} -\bar
V_{K\mu\nu}\frac{\delta}{\delta V_{L\mu\nu}} +\bar
Y^{A}_{K}\frac{\delta}{\delta Y^{A}_{L}} \right)\;,\\
\Bigl\{\mathcal{R}^{(4)}_{IJ},\mathcal{S}_{\Sigma}\Bigr\}&=& \int
d^{4}\!x\,\left(\delta_{IK}\delta_{JL}+\delta_{IL}\delta_{JK}\right)
\left(G^{A}_{K}\frac{\delta}{\delta\bar G^{A}_{L}}
-U_{K\mu\nu}\frac{\delta}{\delta\bar U_{L\mu\nu}}
-X^{A}_{K}\frac{\delta}{\delta\bar X^{A}_{L}}\right)\;.
\end{eqnarray}

\section{Renormalization}
In the previous section we established the full set of Ward
identities fulfilled by the action $\Sigma$.  In this section, we
prove that $\Sigma$ is perturbative renormalizable to all orders.

%%%%%%%%%%%%%%%%%%%%%%%%%%%%%%%%%%%%%%%%%%%%%%%%%%%%%%%%%%%%%%%%%%%%%%%
\subsection{Determination of the most general counterterm}
%%%%%%%%%%%%%%%%%%%%%%%%%%%%%%%%%%%%%%%%%%%%%%%%%%%%%%%%%%%%%%%%%%%%%%%
Let us turn our attention to the characterization of the most
general invariant counterterm $\Sigma_{\mathrm{CT}}$ which can be
freely added to $\Sigma$. According to the algebraic
renormalization procedure \cite{Piguet:1995er}, we require that
the perturbed action $(\Sigma + \epsilon\Sigma_{\mathrm{CT}})$
satisfy the same set of Ward identities,
(\ref{diag_gf}--\ref{rigid}), and constraints of $\Sigma$. The
counterterm $\Sigma_{\mathrm{CT}}$ must be an integrated local
polynomial in the fields and sources with dimension bounded by
four, vanishing ghost number and $\mathcal{Q}_{6}$-charge, obeying
the following constraints
\begin{eqnarray}
\mathcal{S}_{\Sigma}(\Sigma_{\mathrm{CT}})&=&0\;,\label{ct1}\\
\frac{\d\Sigma_{\mathrm{CT}}}{\d b^{i}}&=&0\;,\label{ct2}\\
\frac{\d\Sigma_{\mathrm{CT}}}{\d\bar{c}^{i}}
+\p_{\mu}\frac{\d\Sigma_{\mathrm{CT}}}{\d\Omega^{i}_{\mu}}&=&0\;,\label{ct3}\\
\mathcal{D}_{\S}(\Sigma_{\mathrm{CT}})&=&0\;,\label{ct4}\\
\mathcal{G}^{i}(\Sigma_{\mathrm{CT}})&=&0\;,\label{ct5}\\
\mathcal{W}^{i}(\Sigma_{\mathrm{CT}})&=&0\;,\label{ct6}\\
\mathcal{Q}_{IJ}(\Sigma_{\mathrm{CT}})&=&0\;,\label{ct7}\\
\mathcal{R}^{(\mathcal{N})}_{IJ}(\Sigma_{\mathrm{CT}})&=&0\;.\label{ct8}
\end{eqnarray}
As an immediate consequence of BRST invariance, condition
\eqref{ct1} allows us to write
\begin{equation}
\Sigma_{\mathrm{CT}}=a_{0}\,S_{\mathrm{YM}}+\mathcal{S}_{\Sigma}\Delta^{(-1)}\;,\label{CT}
\end{equation}
where $a_0$ is a free parameter and $\Delta^{(-1)}$ is an
integrated local polynomial with ghost number $-1$ and vanishing
$\mathcal{Q}_{6}$-charge. Taking table (\ref{table1}) into account
and imposing the conditions (\ref{ct2}) and (\ref{ct3}), we are
led to the expression
\begin{equation}
 \Delta^{(-1)} = \Delta^{(-1)}_1 + \Delta^{(-1)}_2 \;,
 \label{new4}
\end{equation}
with
\begin{eqnarray}
\Delta^{(-1)}_1&\!\!\!=\!\!\!&\Int\,\Bigl\{
a_{1}\,\Omega^{a}_{\mu}A^{a}_{\mu}
+a_{2}\,(\p_{\mu}\bar{c}^{a})A^{a}_{\mu}
+a_{3}\,(\Omega^{i}_{\mu}+\p_{\mu}\bar{c}^{i})A^{i}_{\mu}
+a_{4}\,c^{a}L^{a} +a_{5}\,c^{i}L^{i}
+a_{6}\,gf^{abi}\bar{c}^{a}\bar{c}^{b}c^{i}\nonumber\\
&&+a_{7}\,gf^{abc}\bar{c}^{a}\bar{c}^{b}c^{c}
+a_{8}\,\bar{c}^{a}b^{a}
+a_{9}\,gf^{abi}\bar{c}^{a}A^{i}_{\mu}A^{b}_{\mu}
+a_{10}\,\bar{Y}^{A}_{I}B^{A}_{I} +a_{11}\,Y^{A}_{I}\bar{B}^{A}_{I}
+a_{12}\,\bar{X}^{A}_{I}G^{A}_{I}\nonumber\\
&&+a_{13}\,X^{A}_{I}\bar{G}^{A}_{I}
+a_{14}\,\bar{G}^{A}_{I}\p^{2}B^{A}_{I}
+a_{15}\,gf^{ABC}\bar{G}^{A}_{I}(\p_{\mu}A^{B}_{\mu})B^{C}_{I}
+a_{16}\,gf^{ABC}\bar{G}^{A}_{I}A^{B}_{\mu}\p_{\mu}B^{C}_{I}\nonumber\\
&&+a_{17}\,(\p_{\mu}A^{A}_{\nu})\bar{U}_{I\mu\nu}B^{A}_{I}
+a_{18}\,A^{A}_{\mu}\bar{U}_{I\mu\nu}\p_{\nu}B^{A}_{I}
+a_{19}\,gf^{ABC}A^{B}_{\mu}A^{C}_{\nu}\bar{U}_{I\mu\nu}B^{A}_{I}
+a_{20}\,(\p_{\mu}A^{A}_{\nu})V_{I\mu\nu}\bar{G}^{A}_{I}\nonumber\\
&&+a_{21}\,A^{A}_{\mu}V_{I\mu\nu}\p_{\nu}\bar{G}^{A}_{I}
+a_{22}\,gf^{ABC}A^{B}_{\mu}A^{C}_{\nu}V_{I\mu\nu}\bar{G}^{A}_{I}
+a_{23}\,\bar{G}^{A}_{I}B^{A}_{I}\bar{V}_{J\mu\nu}V_{J\mu\nu}
+a_{24}\,\bar{G}^{A}_{I}B^{A}_{J}\bar{V}_{I\mu\nu}V_{J\mu\nu}\nonumber\\
&&+a_{25}\,\bar{G}^{A}_{I}B^{A}_{J}\bar{V}_{J\mu\nu}V_{I\mu\nu}
+a_{26}\,\bar{G}^{A}_{I}B^{A}_{I}\bar{U}_{J\mu\nu}U_{J\mu\nu}
+a_{27}\,\bar{G}^{A}_{I}B^{A}_{J}\bar{U}_{I\mu\nu}U_{J\mu\nu}
+a_{28}\,\bar{G}^{A}_{I}B^{A}_{J}\bar{U}_{J\mu\nu}U_{I\mu\nu}\nonumber\\
&&+a_{29}\bar{G}^{A}_{I}G^{A}_{J}V_{I\mu\nu}\bar{U}_{J\mu\nu}
+a_{30}\bar{G}^{A}_{I}G^{A}_{J}V_{J\mu\nu}\bar{U}_{I\mu\nu}
+a_{31}\bar{B}^{A}_{I}B^{A}_{J}V_{I\mu\nu}\bar{U}_{J\mu\nu}
+a_{32}\bar{B}^{A}_{I}B^{A}_{J}V_{J\mu\nu}\bar{U}_{I\mu\nu}\nonumber\\
&&+a_{33}\,B^{A}_{I}B^{A}_{I}\bar{U}_{J\mu\nu}\bar{V}_{J\mu\nu}
+a_{34}\,B^{A}_{I}B^{A}_{J}\bar{U}_{I\mu\nu}\bar{V}_{J\mu\nu}
+a_{35}\,\bar{G}^{A}_{I}\bar{G}^{A}_{J}{U}_{I\mu\nu}{V}_{J\mu\nu}
+a_{36}\,\bar{G}^{A}_{I}\bar{B}^{A}_{I}{V}_{J\mu\nu}{V}_{J\mu\nu}\nonumber\\
&&+a_{37}\,\bar{G}^{A}_{I}\bar{B}^{A}_{J}{V}_{I\mu\nu}{V}_{J\mu\nu}
+a_{38}\,{G}^{A}_{I}{B}^{A}_{J}\bar{U}_{I\mu\nu}\bar{U}_{J\mu\nu}
+\rho^{ABCD}\,\bar{G}^{A}_{I}B^{B}_{I}A^{C}_{\mu}A^{D}_{\mu}
+a_{39}\,\bar{c}^{a}c^{a}\bar{U}_{I\mu\nu}V_{I\mu\nu}\nonumber\\
&&+\xi_{1}^{abcd}\,\bar{c}^{a}c^{b}\bar{G}^{c}_{I}B^{d}_{I}
+\xi_{2}^{abij}\,\bar{c}^{a}c^{b}\bar{G}^{i}_{I}B^{j}_{J}
+\s_{1}^{ABCD}\,\bar{G}^{A}_{I}B^{B}_{I}\bar{B}^{C}_{J}B^{D}_{J}
+\s_{2}^{ABCD}\,\bar{G}^{A}_{I}B^{B}_{I}\bar{G}^{C}_{J}G^{D}_{J}\nonumber\\
&&+\s_{3}^{ABCD}\,\bar{G}^{A}_{I}B^{B}_{J}\bar{B}^{C}_{I}B^{D}_{J}
+\s_{4}^{ABCD}\,\bar{G}^{A}_{I}B^{B}_{J}\bar{G}^{C}_{I}G^{D}_{J}\Bigr\}
\;,\label{delta1}
\end{eqnarray}
and
\begin{eqnarray}
\Delta^{(-1)}_{2}&=&\Int\,\Bigl(
a_{40}\,\bar{U}_{I\mu\nu}\p^{2}V_{I\mu\nu}
+a_{41}\,\bar{U}_{I\mu\nu}\p_{\mu}\p_{\alpha}V_{I\nu\alpha}
+a_{42}\,\bar{U}_{I\mu\nu}V_{I\mu\nu}\bar{U}_{J\alpha\beta}U_{J\alpha\beta}\nonumber\\
&&+a_{43}\,\bar{U}_{I\mu\nu}V_{I\mu\nu}\bar{V}_{J\alpha\beta}V_{J\alpha\beta}
+a_{44}\,\bar{U}_{I\mu\nu}V_{J\mu\nu}\bar{U}_{I\alpha\beta}U_{J\alpha\beta}
+a_{45}\,\bar{U}_{I\mu\nu}V_{J\mu\nu}\bar{V}_{I\alpha\beta}V_{J\alpha\beta}\nonumber\\
&&+a_{46}\,\bar{U}_{I\mu\nu}V_{I\nu\alpha}\bar{U}_{J\alpha\beta}U_{J\beta\mu}
+a_{47}\,\bar{U}_{I\mu\nu}V_{I\nu\alpha}\bar{V}_{J\alpha\beta}V_{J\beta\mu}
+a_{48}\,\bar{U}_{I\mu\nu}V_{J\nu\alpha}\bar{U}_{I\alpha\beta}U_{J\beta\mu}\nonumber\\
&&+a_{49}\,\bar{U}_{I\mu\nu}V_{J\nu\alpha}\bar{V}_{I\alpha\beta}V_{J\beta\mu}
+a_{50}\,{V}_{I\mu\nu}V_{I\mu\nu}\bar{U}_{J\alpha\beta}\bar{V}_{J\alpha\beta}
+a_{51}\,{V}_{I\mu\nu}V_{J\mu\nu}\bar{U}_{I\alpha\beta}\bar{V}_{J\alpha\beta}\nonumber\\
&&+a_{52}\,\bar{U}_{I\mu\nu}\bar{U}_{J\mu\nu}{V}_{J\alpha\beta}{U}_{J\alpha\beta}
+a_{53}\,\bar{U}_{I\mu\nu}V_{I\nu\alpha}\bar{U}_{J\mu\beta}U_{J\alpha\beta}
+a_{54}\,\bar{U}_{I\mu\nu}V_{I\nu\alpha}\bar{V}_{J\mu\beta}V_{J\alpha\beta}\nonumber\\
&&+a_{55}\,\bar{U}_{I\mu\nu}V_{J\nu\alpha}\bar{U}_{I\mu\beta}U_{J\alpha\beta}
+a_{56}\,\bar{U}_{I\mu\nu}V_{J\nu\alpha}\bar{V}_{I\mu\beta}V_{J\alpha\beta}
+a_{57}\,\bar{U}_{I\mu\nu}V_{I\alpha\beta}\bar{U}_{J\alpha\beta}U_{J\mu\nu}\nonumber\\
&&+a_{58}\,\bar{U}_{I\mu\nu}V_{I\alpha\beta}\bar{V}_{J\alpha\beta}V_{J\mu\nu}
+a_{59}\,\bar{U}_{I\alpha\mu}V_{I\nu\beta}\bar{U}_{J\alpha\beta}U_{J\mu\nu}
+a_{60}\,\bar{U}_{I\alpha\mu}V_{I\nu\beta}\bar{V}_{J\alpha\beta}V_{J\mu\nu}\Bigr)
\label{delta2}
\end{eqnarray}
Here $a_\alpha$, $\alpha = 1, \dots, 60$,  and the four-rank
tensor $\sigma_\aleph^{ABCD}$, $\aleph = 1,\dots,4$, are also
arbitrary coefficients.  Applying the remaining conditions
(\ref{ct4}--\ref{ct8}), expression \eqref{new4} reduces to
\begin{eqnarray}
\Delta^{(-1)}&\!\!\!=\!\!\!&\Int\Bigl\{a_{1}\,\Omega^{a}_{\mu}A^{a}_{\mu}
-a_{2}\,\bar{c}^{a}D^{ab}_{\mu}A^{b}_{\mu} +a_{4}\,c^{a}L^{a}
+a_{8}\,\left(\bar{c}^{a}b^{a} -gf^{abi}\bar{c}^{a}\bar{c}^{b}c^{i}
-\frac{g}{2}f^{abc}\bar{c}^{a}\bar{c}^{b}c^{c}\right)\nonumber\\
&&+a_{10}\,\left(\bar{Y}^{A}_{I}B^{A}_{I}
-\bar{X}^{A}_{I}G^{A}_{I}\right)
+a_{11}\,\left(Y^{A}_{I}\bar{B}^{A}_{I}
-X^{A}_{I}\bar{G}^{A}_{I}\right)
+a_{14}\,\bar{G}^{A}_{I}D^{AC}_{\mu}D^{CB}_{\mu}B^{B}_{I}\nonumber\\
&&+a_{19}\,F^{A}_{\mu\nu}\bar{U}_{I\mu\nu}B^{A}_{I}
+(a_{10}-a_{11}-a_{19})\,F^{A}_{\mu\nu}V_{I\mu\nu}\bar{G}^{A}_{I}
+a_{23}\,\bar{G}^{A}_{I}B^{A}_{I}\left(\bar{V}_{J\mu\nu}V_{J\mu\nu}
-\bar{U}_{J\mu\nu}U_{J\mu\nu}\right)\nonumber\\
&&+a_{25}\,\left(\bar{G}^{A}_{I}B^{A}_{J}\bar{V}_{J\mu\nu}V_{I\mu\nu}
+ \bar{G}^{A}_{I}G^{A}_{J}V_{I\mu\nu}\bar{U}_{J\mu\nu}\right)
+a_{28}\,\left(\bar{G}^{A}_{I}B^{A}_{J}\bar{U}_{J\mu\nu}U_{I\mu\nu}
+\bar{B}^{A}_{I}B^{A}_{J}V_{I\mu\nu}\bar{U}_{J\mu\nu}\right)\nonumber\\
&&+\half[\lambda_{2}(a_{10}-a_{11})-a_{28}-a_{25}]
\,\left(B^{A}_{I}B^{A}_{J}\bar{U}_{I\mu\nu}\bar{V}_{J\mu\nu} +
G^{A}_{I}B^{A}_{J}\bar{U}_{I\mu\nu}\bar{U}_{J\mu\nu}\right)\nonumber\\
&&-\half[\lambda_{2}(a_{10}-a_{11})+a_{28}+a_{25}]
\,\left(\bar{G}^{A}_{I}\bar{G}^{A}_{J}U_{I\mu\nu}V_{J\mu\nu}
+\bar{G}^{A}_{I}\bar{B}^{A}_{J}V_{I\mu\nu}V_{J\mu\nu}\right)\nonumber\\
&&+\sigma^{ABCD}_{1}\,\bar{G}^{A}_{I}B^{B}_{I}\left(\bar{B}^{C}_{J}B^{D}_{J}
-\bar{G}^{C}_{J}G^{D}_{J}\right)
+a_{40}\,\bar{U}_{I\mu\nu}\p^{2}V_{I\mu\nu}
+a_{41}\,\bar{U}_{I\mu\nu}\p_{\mu}\p_{\alpha}V_{I\nu\alpha}\nonumber\\
&&+a_{42}\,\left(\bar{U}_{I\mu\nu}V_{I\mu\nu}\bar{U}_{J\alpha\beta}U_{J\alpha\beta}
-\bar{U}_{I\mu\nu}V_{I\mu\nu}\bar{V}_{J\alpha\beta}V_{J\alpha\beta}\right)
\Bigr\}\;,\label{final_delta}
\end{eqnarray}
with the further restrictions on $\s_{1}^{ABCD}$
\begin{equation}
f^{MAN}\s_{1}^{MBCD}+ f^{MBN}\s_{1}^{AMCD} +f^{MCN}\s_{1}^{ABMD}
+f^{MDN}\s_{1}^{ABCM}=0\;,
\end{equation}
and
\begin{equation}
\s_{1}^{ABCD}=\s_{1}^{CDAB}=\s_{1}^{BACD}\;.
\end{equation}
Observe that many coefficients present in expressions
\eqref{delta1} and \eqref{delta2} vanish after the imposition of
the constraints (\ref{ct4}--\ref{ct8}). \\\\Performing now the
following redefinition
\begin{eqnarray}
a_{0}&\to&a_{0}\;,\nonumber\\
a_{1}&\to&a_{1}\;,\nonumber\\
a_{2}&\to&-a_{2}\;,\nonumber\\
a_{4}&\to&-a_{3}\;,\nonumber\\
a_{8}&\to&\frac{\alpha}{2}a_{4}\;,\nonumber\\
a_{10}+a_{11}&\to&a_{5}\;,\nonumber\\
a_{10}-a_{19}&\to&a_{6}\;,\nonumber\\
a_{23}&\to&\lambda_{1}a_{7}\;,\nonumber\\
a_{25}+a_{28}&\to&\lambda_{2}a_{8}\;,\nonumber\\
a_{40}&\to&\chi_{1}a_{9}\;,\nonumber\\
a_{41}&\to&\chi_{2}a_{10}\;,\nonumber\\
a_{42}&\to&\zeta a_{11}\;,\nonumber\\
\sigma_{1}^{ABCD}&\to&\frac{a_{5}}{16}(\sigma^{ABCD}-\lambda^{ABCD})\;,
\end{eqnarray}
we can rewrite the counterterm  $\Sigma_{\mathrm{CT}}$ as
\begin{equation}
 \Sigma_{\mathrm{CT}}=\Sigma_{\mathrm{CT}(1)} + \Sigma_{\mathrm{CT}(2)}
 + \Sigma_{\mathrm{CT}(3)}
\label{final_CT}
\end{equation}
with
\begin{eqnarray}
\Sigma_{\mathrm{CT}(1)}&\!\!=\!\!&\Int\,\biggl\{
\frac{a_{0}+2a_{1}}{2}\left(D^{ab}_{\mu}A^{b}_{\nu}\right)
\left(D^{ac}_{\mu}A^{c}_{\nu} -D^{ac}_{\nu}A^{c}_{\mu}\right)
+(a_{0}+3a_{1})gf^{abc}\left(D^{ad}_{\mu}A^{d}_{\nu}\right)
A^{b}_{\mu}A^{c}_{\nu}\nonumber\\
&&+\frac{a_{0}+4a_{1}}{4}\left(
g^{2}f^{abc}f^{ade}A^{b}_{\mu}A^{c}_{\nu}A^{d}_{\mu}A^{e}_{\nu}
+g^{2}f^{abi}f^{cdi}A^{a}_{\mu}A^{b}_{\nu}A^{c}_{\mu}A^{d}_{\nu}\right)
+\frac{a_{0}}{2}\left(\p_{\mu}A^{i}_{\nu}\right)
\left(\p_{\mu}A^{i}_{\nu}-\p_{\nu}A^{i}_{\mu}\right)\nonumber\\
&&+(a_{0}+2a_{1})gf^{abi}\left(\p_{\mu}A^{i}_{\nu}\right)
A^{a}_{\mu}A^{b}_{\nu} +(a_{1}+a_{2})b^{a}D^{ab}_{\mu}A^{a}_{\mu}
+(a_{1}+a_{2})gf^{abi}\bar{c}^{a}
\left(D^{bc}_{\mu}A^{c}_{\mu}\right)c^{i}\nonumber\\
&&+(a_{1}+a_{2}+a_{3})\bar{c}^{a}D^{ab}_{\mu}
\left(gf^{bcd}A^{c}_{\mu}c^{d}\right)
+(2a_{1}+a_{2}+a_{3})g^{2}f^{abi}f^{cdi}\bar{c}^{a}c^{d}A^{b}_{\mu}A^{c}_{\mu}\nonumber\\
&&-(a_{1}+a_{3})\!\left(\Omega^{i}_{\mu}+\p_{\mu}\bar{c}^{i}\right)
gf^{abi}A^{a}_{\mu}c^{b}
+(a_{2}+a_{3})\bar{c}^{a}D^{ab}_{\mu}D^{bc}_{\mu}c^{c}
+a_{4}\frac{\alpha}{2}b^{a}b^{a} -a_{4}\alpha
gf^{abi}b^{a}\bar{c}^{b}c^{i}\nonumber\\
&&-(a_{3}+a_{4})\!\left[\frac{\alpha}{2}gf^{abc}b^{a}\bar{c}^{b}c^{c}
+\frac{\alpha}{4}g^{2}f^{abc}f^{adi}\bar{c}^{b}\bar{c}^{c}c^{d}c^{i}\right]
-(2a_{3}+a_{4})
\!\left[\frac{\alpha}{4}g^{2}f^{abi}f^{cdi}\bar{c}^{a}\bar{c}^{b}c^{c}c^{d}\right.\nonumber\\
&&+\left.\!\frac{\alpha}{8}g^{2}f^{abc}f^{ade}\bar{c}^{b}\bar{c}^{c}c^{d}c^{e}\right]
+(a_{1}-a_{3})\,\Omega^{a}_{\mu}D^{ab}_{\mu}c^{b}
-a_{3}\,gf^{abc}\Omega^{a}_{\mu}A^{b}_{\mu}c^{c}
+a_{3}\,\frac{g}{2}f^{abc}L^{a}c^{b}c^{c}\nonumber\\
&&+a_{3}\,gf^{abi}L^{i}c^{a}c^{b}
\biggl\}\;,\\
\Sigma_{\mathrm{CT}(2)}&\!\!=\!\!&\Int\,\biggl\{
a_{3}\,\bar{Y}^{a}_{I}\left(gf^{abc}c^{b}B^{c}_{I}
+gf^{abi}c^{b}B^{i}_{I}\right)
+a_{3}\,f^{abi}\bar{Y}^{i}_{I}c^{a}B^{b}_{I}
+a_{3}\,Y^{a}_{I}\left(gf^{abc}c^{b}\bar{B}^{c}_{I}\right.\nonumber\\
&&+\left.\!gf^{abi}c^{b}\bar{B}^{i}_{I}\right)
+a_{3}\,gf^{abi}Y^{i}_{I}c^{a}\bar{B}^{b}_{I}
+a_{3}\,\bar{X}^{a}_{I}\left(gf^{abc}c^{b}G^{c}_{I}
+gf^{abi}c^{b}G^{i}_{I}\right)
+a_{3}\,gf^{abi}\bar{X}^{i}_{I}c^{a}G^{b}_{I}\nonumber\\
&&+a_{3}\,X^{a}_{I}\left(gf^{abc}c^{b}\bar{G}^{c}_{I}
+gf^{abi}c^{b}\bar{G}^{i}_{I}\right)
+a_{3}\,gf^{abi}X^{i}_{I}c^{a}\bar{G}^{b}_{I}
+a_{5}\,\bar{B}^{a}_{I}\p^{2}B^{a}_{I}
-a_{5}\,gf^{abi}\bar{B}^{a}_{I}\left(\p_{\mu}A^{i}_{\mu}\right)B^{b}_{I}\nonumber\\
&&-(a_{1}+a_{5})gf^{abc}\bar{B}^{a}_{I}\left(\p_{\mu}A^{c}_{\mu}\right)B^{b}_{I}
-a_{5}\,2gf^{abi}\bar{B}^{a}_{I}A^{i}_{\mu}\p_{\mu}B^{b}_{I}
-(a_{1}+a_{5})2gf^{abc}\bar{B}^{a}_{I}A^{c}_{\mu}\p_{\mu}B^{b}_{I}\nonumber\\
&&+a_{5}\,g^{2}f^{aci}f^{cbj}\bar{B}^{a}_{I}A^{i}_{\mu}A^{j}_{\mu}B^{b}_{I}
+(a_{1}+a_{5})g^{2}f^{aci}f^{cbd}\bar{B}^{a}_{I}A^{i}_{\mu}A^{d}_{\mu}B^{b}_{I}
+(a_{1}+a_{5})g^{2}f^{acd}f^{cdi}\bar{B}^{a}_{I}A^{d}_{\mu}A^{i}_{\mu}B^{b}_{I}\nonumber\\
&&+(2a_{1}+a_{5})g^{2}f^{acd}f^{cbe}\bar{B}^{a}_{I}A^{d}_{\mu}A^{e}_{\mu}B^{b}_{I}
+(2a_{1}+a_{5})g^{2}f^{aci}f^{bdi}\bar{B}^{a}_{I}A^{c}_{\mu}A^{d}_{\mu}B^{b}_{I}
+a_{5}\,\bar{B}^{i}_{I}\p^{2}B^{i}_{I}\nonumber\\
&&-(2a_{1}+a_{5})g^{2}f^{abi}f^{acj}\bar{B}^{i}_{I}A^{b}_{\mu}A^{c}_{\mu}B^{j}_{I}
+(a_{1}+a_{5})gf^{abi}\bar{B}^{a}_{I}\left(\p_{\mu}A^{b}_{\mu}\right)B^{i}_{I}
+(a_{1}+a_{5})2gf^{abi}\bar{B}^{a}_{I}A^{b}_{\mu}\p_{\mu}B^{i}_{I}\nonumber\\
&&-(a_{1}+a_{5})g^{2}f^{acj}f^{cbi}\bar{B}^{a}_{I}A^{j}_{\mu}A^{b}_{\mu}B^{i}_{I}
-(2a_{1}+a_{5})g^{2}f^{acb}f^{cdi}\bar{B}^{a}_{I}A^{b}_{\mu}A^{d}_{\mu}B^{i}_{I}\nonumber\\
&&-(a_{1}+a_{5})gf^{abi}\bar{B}^{i}\left(\p_{\mu}A^{b}_{\mu}\right)B^{a}_{I}
-(a_{1}+a_{5})2gf^{abi}\bar{B}^{i}_{I}A^{b}_{\mu}\p_{\mu}B^{a}_{I}
+(a_{1}+a_{5})g^{2}f^{cbi}f^{caj}\bar{B}^{i}_{I}A^{b}_{\mu}A^{j}_{\mu}B^{a}_{I}\nonumber\\
&&+(2a_{1}+a_{5})g^{2}f^{cbi}f^{cad}\bar{B}^{i}_{I}A^{b}_{\mu}A^{d}_{\mu}B^{a}_{I}
\biggl\} \;,
\end{eqnarray}
and
\begin{eqnarray}
\Sigma_{\mathrm{CT}(3)}&\!\!=\!\!&\Int\,\biggl\{
-\Bigl[a_{5}\,\bar{G}^{a}_{I}\p^{2}G^{a}_{I}
-a_{5}\,gf^{abi}\bar{G}^{a}_{I}\left(\p_{\mu}A^{i}_{\mu}\right)G^{b}_{I}\nonumber\\
&&-(a_{1}+a_{5})gf^{abc}\bar{G}^{a}_{I}\left(\p_{\mu}A^{c}_{\mu}\right)G^{b}_{I}
-a_{5}\,2gf^{abi}\bar{G}^{a}_{I}A^{i}_{\mu}\p_{\mu}G^{b}_{I}
-(a_{1}+a_{5})2gf^{abc}\bar{G}^{a}_{I}A^{c}_{\mu}\p_{\mu}G^{b}_{I}\nonumber\\
&&+a_{5}\,g^{2}f^{aci}f^{cbj}\bar{G}^{a}_{I}A^{i}_{\mu}A^{j}_{\mu}G^{b}_{I}
+(a_{1}+a_{5})g^{2}f^{aci}f^{cbd}\bar{G}^{a}_{I}A^{i}_{\mu}A^{d}_{\mu}G^{b}_{I}
+(a_{1}+a_{5})g^{2}f^{acd}f^{cdi}\bar{G}^{a}_{I}A^{d}_{\mu}A^{i}_{\mu}G^{b}_{I}\nonumber\\
&&+(2a_{1}+a_{5})g^{2}f^{acd}f^{cbe}\bar{G}^{a}_{I}A^{d}_{\mu}A^{e}_{\mu}G^{b}_{I}
+(2a_{1}+a_{5})g^{2}f^{aci}f^{bdi}\bar{G}^{a}_{I}A^{c}_{\mu}A^{d}_{\mu}G^{b}_{I}
+a_{5}\,\bar{G}^{i}_{I}\p^{2}G^{i}_{I}\nonumber\\
&&-(2a_{1}+a_{5})g^{2}f^{abi}f^{acj}\bar{G}^{i}_{I}A^{b}_{\mu}A^{c}_{\mu}G^{j}_{I}
+(a_{1}+a_{5})gf^{abi}\bar{G}^{a}_{I}\left(\p_{\mu}A^{b}_{\mu}\right)G^{i}_{I}
+(a_{1}+a_{5})2gf^{abi}\bar{G}^{a}_{I}A^{b}_{\mu}\p_{\mu}G^{i}_{I}\nonumber\\
&&-(a_{1}+a_{5})g^{2}f^{acj}f^{cbi}\bar{G}^{a}_{I}A^{j}_{\mu}A^{b}_{\mu}G^{i}_{I}
-(2a_{1}+a_{5})g^{2}f^{acb}f^{cdi}\bar{G}^{a}_{I}A^{b}_{\mu}A^{d}_{\mu}G^{i}_{I}\nonumber\\
&&-(a_{1}+a_{5})gf^{abi}\bar{G}^{i}\left(\p_{\mu}A^{b}_{\mu}\right)G^{a}_{I}
-(a_{1}+a_{5})2gf^{abi}\bar{G}^{i}_{I}A^{b}_{\mu}\p_{\mu}G^{a}_{I}
+(a_{1}+a_{5})g^{2}f^{cbi}f^{caj}\bar{G}^{i}_{I}A^{b}_{\mu}A^{j}_{\mu}G^{a}_{I}\nonumber\\
&&+(2a_{1}+a_{5})g^{2}f^{cbi}f^{cad}\bar{G}^{i}_{I}A^{b}_{\mu}A^{d}_{\mu}G^{a}_{I}\Bigr]
+\left[2(a_{1}+a_{6})\,D^{ab}_{\mu}A^{b}_{\nu}
+(2a_{1}+a_{6})gf^{abc}A^{b}_{\mu}A^{c}_{\nu}\right]
\left(\bar{U}_{I\mu\nu}\,G^{a}_{I}\right.\nonumber\\
&&+\left.\!V_{I\mu\nu}\,\bar{B}^{a}_{I}
-\bar{V}_{I\mu\nu}\,B^{a}_{I} +U_{I\mu\nu}\,\bar{G}^{a}_{I}\right)
+\left[2a_{6}\,\p_{\mu}A^{i}_{\nu}
+(2a_{1}+a_{6})gf^{abi}A^{a}_{\mu}A^{b}_{\nu}\right]
\left(\bar{U}_{I\mu\nu}\,G^{i}_{I}\right.\nonumber\\
&&+\left.\!V_{I\mu\nu}\,\bar{B}^{i}_{I}
-\bar{V}_{I\mu\nu}\,B^{i}_{I} +U_{I\mu\nu}\,\bar{G}^{i}_{I}\right)
+\lambda_{1}(a_{5}+a_{7})\left(\bar{B}^{A}_{I}B^{A}_{I}
-\bar{G}^{A}_{I}G^{A}_{I}\right) \left(\bar{V}_{J\mu\nu}V_{J\mu\nu}
-\bar{U}_{J\mu\nu}U_{J\mu\nu}\right) \nonumber\\&&
 +\lambda_{2}(a_{5}+a_{8})
\Bigl(\bar{B}^{A}_{I}B^{A}_{J}V_{I\mu\nu}\bar{V}_{J\mu\nu}
-\bar{G}^{A}_{I}B^{A}_{J}U_{I\mu\nu}\bar{V}_{J\mu\nu}
-G^{A}_{I}B^{A}_{J}\bar{U}_{I\mu\nu}\bar{V}_{J\mu\nu}\nonumber\\
&&+\bar{B}^{A}_{I}G^{A}_{J}V_{I\mu\nu}\bar{U}_{J\mu\nu}
+\bar{G}^{A}_{I}G^{A}_{J}U_{I\mu\nu}\bar{U}_{J\mu\nu}
+\bar{G}^{A}_{I}\bar{B}^{A}_{J}U_{I\mu\nu}V_{J\mu\nu} -\half
B^{A}_{I}B^{A}_{J}\bar{V}_{I\mu\nu}\bar{V}_{J\mu\nu} +\half
G^{A}_{I}G^{A}_{J}\bar{U}_{I\mu\nu}\bar{U}_{J\mu\nu}\nonumber\\
&&-\half \bar{B}^{A}_{I}\bar{B}^{A}_{J}V_{I\mu\nu}V_{J\mu\nu} +\half
\bar{G}^{A}_{I}\bar{G}^{A}_{J}U_{I\mu\nu}U_{J\mu\nu} \Bigr)
+\frac{a_{5}}{16}(\lambda^{ABCD}+\sigma^{ABCD})
(\bar{B}^{A}_{I}B^{B}_{I}-\bar{G}^{A}_{I}G^{B}_{I})
(\bar{B}^{C}_{J}B^{D}_{J}-\bar{G}^{C}_{J}G^{D}_{J})\nonumber\\
&&+\chi_{1}a_{9}(\bar{V}_{I\mu\nu}\p^{2}V_{I\mu\nu}
-\bar{U}_{I\mu\nu}\p^{2}U_{I\mu\nu})
+\chi_{2}a_{10}(\bar{V}_{I\mu\nu}\p_{\mu}\p_{\alpha}V_{I\nu\alpha}
-\bar{U}_{I\mu\nu}\p_{\mu}\p_{\alpha}U_{I\nu\alpha})\nonumber\\
&&-\zeta a_{11}(\bar{U}_{I\mu\nu}U_{I\mu\nu}
\bar{U}_{J\alpha\beta}U_{J\alpha\beta}
+\bar{V}_{I\mu\nu}V_{I\mu\nu}
\bar{V}_{J\alpha\beta}V_{J\alpha\beta}
-2\bar{V}_{I\mu\nu}V_{I\mu\nu}
\bar{U}_{J\alpha\beta}U_{J\alpha\beta})\biggl\}\;. \label{new5}
\end{eqnarray}
By construction, expression (\ref{final_CT}) yields the most
general invariant counterterm compatible with the full set of Ward
identities.
%%%%%%%%%%%%%%%%%%%%%%%%%%%%%%%%%
\subsection{Renormalization factors}
%%%%%%%%%%%%%%%%%%%%%%%%%%%%%%%%%
Once we have found the most general counterterm,
eq.(\ref{final_CT}), we have to check whether the remaining
independent coefficients, $a_{0},a_{1},\dots,a_{11}$, and the
4-rank tensor $\sigma^{ABCD}$ can be reabsorbed through a
redefinition of the fields, sources and parameters of the starting
action $\Sigma$. The answer is in fact affirmative. Let us rename
collectively the fields, sources and parameters as:
\begin{equation}
\Phi=(A,b,\bar{c},c)\qquad\mathrm{and}\qquad\Psi=(\bar{B},B,\bar{G},G)\;,
\label{fields}
\end{equation}
\begin{equation}
J=(\Omega,L)
\qquad\mathrm{and}\qquad\mathcal{J}=(\bar{Y},Y,\bar{X},X,\bar{U},U,\bar{V},V)\;,
\label{sources-ren}
\end{equation}
\begin{equation}
\xi=(g,\alpha,\lambda_{1},\lambda_{2},\chi_{1},\chi_{2},\zeta)\;.
\label{parameters}
\end{equation}
Next, defining the bare fields, sources and parameters as:
\begin{eqnarray}
\Phi_{0}^{\mbox{\footnotesize{off-diag}}}&=&
\widetilde{Z}^{1/2}_{\Phi}\,\Phi^{\mbox{\footnotesize{off-diag}}}\;,\nonumber\\
\Phi^{\mbox{\footnotesize{diag}}}_{0}
&=&Z_{\Phi}^{1/2}\,\Phi^{\mbox{\footnotesize{diag}}}\;,\nonumber\\
\Psi_{0}&=&Z_{\Psi}^{1/2}\,\Psi\;,\nonumber\\
J_{0}^{\,\mbox{\footnotesize{off-diag}}}
&=&\widetilde{Z}_{J}\,J^{\,\mbox{\footnotesize{off-diag}}}\;,\nonumber\\
J_{0}^{\,\mbox{\footnotesize{diag}}}&=&Z_{J}\,J^{\,\mbox{\footnotesize{diag}}}\;,\nonumber\\
\mathcal{J}_{0}&=&Z_{\mathcal{J}}\,\mathcal{J}\;,\nonumber\\
\xi_{0}&=&Z_{\xi}\,\xi\;,
\label{redefinedfields}
\end{eqnarray}
and
\begin{equation}
 \lambda^{ABCD}_{0}=Z_{\lambda}\,\lambda^{ABCD}+\mathcal{Z}^{ABCD}\;,
 \label{redefinedlambda}
\end{equation}
it is easily checked that the invariant counterterm
$\Sigma_{\mathrm{CT}}$ can be reabsorbed into the starting
classical action $\Sigma$, namely
\begin{equation}
\Sigma[\xi,\Phi,\Psi,J,\mathcal{J},
\lambda^{ABCD}]+\epsilon\,\Sigma_{\mathrm{CT}}=
\Sigma[\xi_{0},\Phi_{0},\Psi_{0},J_{0},\mathcal{J}_{0},\lambda^{ABCD}_{0}]+O(\epsilon^{2})\;,
\label{renor}
\end{equation}
where $\epsilon$ stands for an infinitesimal expansion parameter.
For the $Z$'s factors we have
\begin{eqnarray}
Z_{A}^{1/2}&\!\!\!=\!\!\!&Z_{g}^{-1}\;,\nonumber\\
\widetilde{Z}^{1/2}_{b}&\!\!\!=\!\!\!&Z_{g}Z^{1/2}_{c}\widetilde{Z}^{1/2}_{c}\;,\nonumber\\
Z^{1/2}_{b}&\!\!\!=\!\!\!&Z_{g}\;,\nonumber\\
\widetilde{Z}^{1/2}_{\bar{c}}&\!\!\!=\!\!\!&\widetilde{Z}^{1/2}_{c}\;,\nonumber\\
Z^{1/2}_{\bar{c}}&\!\!\!=\!\!\!&Z^{-1/2}_{\bar{c}}\;,\nonumber\\
Z^{1/2}_{\bar{B}}&\!\!\!=\!\!\!&Z^{1/2}_{\bar{G}}=\;Z^{1/2}_{G}=\;Z^{1/2}_{B}\;,\nonumber\\
\widetilde{Z}_{\Omega}
&\!\!\!=\!\!\!&Z_{g}^{-1}Z^{-1/2}_{c}\widetilde{Z}^{-1/2}_{A}\;,\nonumber\\
Z_{\Omega}&\!\!\!=\!\!\!&Z_{c}^{-1/2}\;,\nonumber\\
\widetilde{Z}_{L}
&\!\!\!=\!\!\!&Z^{-1}_{g}Z^{-1/2}_{c}\widetilde{Z}^{-1/2}_{c}\;,\nonumber\\
Z_{L}&\!\!\!=\!\!\!&Z^{-1}_{g}Z^{-1}_{c}\;,\nonumber\\
Z_{X}&\!\!\!=\!\!\!&Z_{\bar{X}}=\;Z_{Y}=\;Z_{\bar{Y}}
=\;Z^{-1}_{g}Z^{-1/2}_{c}Z^{-1/2}_{B}\;,\nonumber\\
Z_{U}&\!\!\!=\!\!\!&Z_{\bar{U}}=\;Z_{\bar{V}}=\;Z_{V}\;,
\end{eqnarray}
\noindent with,
\begin{eqnarray}
\widetilde{Z}^{1/2}_{A}&=&1+\epsilon\,\left(\frac{a_{0}}{2}+a_{1}\right)\;,\nonumber\\
Z_{g}&=&1-\epsilon\,\frac{a_{0}}{2}\;,\nonumber\\
\widetilde{Z}^{1/2}_{c}&=&1+\epsilon\,\frac{a_{2}+a_{3}}{2}\;,\nonumber\\
Z^{1/2}_{c}&=&1+\epsilon\,\frac{a_{2}-a_{3}}{2}\;,\nonumber\\
Z^{1/2}_{B}&=&1+\epsilon\,\frac{a_{5}}{2}\;,\nonumber\\
Z_{V}&=&1-\epsilon\,\left(\frac{a_{0}}{2}+\frac{a_{5}}{2}-a_{6}\right)\;,\nonumber\\
Z_{\alpha}&=&1+\epsilon\,(a_{0}-2a_{2}+a_{4})\;,\nonumber\\
Z_{\lambda_{1}}&=&1+\epsilon\,(a_{0}+a_{5}-2a_{6}+a_{7})\;,\nonumber\\
Z_{\lambda_{2}}&=&1+\epsilon\,(a_{0}+a_{5}-2a_{6}+a_{8})\;,\nonumber\\
Z_{\chi_{1}}&=&1+\epsilon\,(a_{0}+a_{5}-2a_{6}+a_{9})\;,\nonumber\\
Z_{\chi_{2}}&=&1+\epsilon\,(a_{0}+a_{5}-2a_{6}+a_{10})\;,\nonumber\\
Z_{\zeta}&=&1+\epsilon\,(2a_{0}+2a_{5}-4a_{6}+a_{11})\;,\nonumber\\
Z_{\lambda}&=&1-\epsilon\,a_{5}\;,\nonumber\\
\mathcal{Z}^{ABCD}&=&\epsilon\,a_{5}\sigma^{ABCD}\;.
\end{eqnarray}
This concludes the proof of the renormalizability of the classical
action to all orders of perturbation theory.

%%%%%%%%%%%%%%%%%%
\section{Conclusions}

%In this paper we have proven that the introduction of the nonlocal
 %gauge invariant mass operator $\mathrm{Tr} \int d^4x
%F_{\mu\nu} (D^2)^{-1} F_{\mu\nu}$ to the YM action fixed in the
%maximal Abelian gauge leads to a
% multiplicative renormalizable theory.

In this paper, a detailed analysis of the nonlocal gauge invariant
mass operator $Tr \int d^4x F_{\mu\nu} (D^2)^{-1} F_{\mu\nu}$ has
been made in the MAG. By means of the introduction of a suitable
set of auxiliary fields, this operator can be cast in local form.
Moreover, the embedding of the resulting local model into a more
general action  has allowed us to make use of the BRST symmetry.
Furthermore, it turns out that the generalized action displays
additional global symmetries giving rise to useful Ward
identities, which were used to restrict the possible counterterms.
The analysis of the renormalization factors has enabled us to show
that the most general invariant counterterm can be in fact
reabsorbed into the starting action through a redefinition of
fields. parameter and sources, establishing thus the perturbative
renormalizability of the model to all orders. Finally, in appendix
{\bf A} the nonlocal operator $Tr \int d^4x F_{\mu\nu} (D^2)^{-1}
F_{\mu\nu}$ has been analyzed in the presence of the horizon
function implementing the restriction of the domain of integration
in the Feynman path integral to the Gribov region in the MAG. The
output of our analysis is that the introduction of the horizon
function does not spoil the renormalizability of the model.

\section*{Acknowledgments}

The Conselho Nacional de Desenvolvimento Cient\'{i}fico e
Tecnol\'{o}gico (CNPq-Brazil), the SR2-UERJ and the
Coordena{\c{c}}{\~{a}}o de Aperfei{\c{c}}oamento de Pessoal de
N{\'{i}}vel Superior (CAPES) are gratefully acknowledged for
financial support.

\appendix

\section{Including the horizon function}
It is a well known fact that non-Abelian theories are plagued by
Gribov ambiguities \cite{Gribov:1977wm}, see
\cite{Sobreiro:2005ec} for a pedagogical review. In the specific
case of the MAG, the study of Gribov ambiguities and the
characterization of the horizon function in the case of $SU(2)$
can be found in \cite{Capri:2005tj,Capri:2006cz}. In this appendix
we present the main aspects of the simultaneous inclusion of the
MAG horizon function and of the gauge invariant mass operator
\eqref{trmassop} in the YM theory. A similar treatment regarding
Gribov ambiguities and the mass operator \eqref{trmassop} has been
done recently in the Landau gauge \cite{Capri:2007ix}. Without
loss of generality, we will follow
\cite{Capri:2005tj,Capri:2006cz} and restrict ourselves to
$SU(2)$. \\\\In \cite{Capri:2006cz}, the following horizon
function for the MAG has been derived
\begin{equation}
S_{\mathrm{Horizon}}=\gamma^{4}g^{2}\int d^{4}\!x\,
\varepsilon^{ab}A_{\mu}\left({\cal M}^{-1}\right)^{ac}
\varepsilon^{cb}A_{\mu}\;.\label{horizon}
\end{equation}
Here $\gamma$ stands for the Gribov parameter
\cite{Gribov:1977wm,Capri:2005tj}, $\varepsilon^{ab}\equiv
\varepsilon^{3ab},\; a,b=1,2$ are the off-diagonal components of
the $SU(2)$ structure constants and $A_{\mu}\equiv A^{3}_{\mu}$ is
the diagonal component of the gauge field. The operator
$\left({\cal M}^{-1}\right)^{ab}$ is the inverse of the
Faddeev-Popov operator given by
\begin{equation}
\mathcal{M}^{ab}=-D^{ac}_{\mu}D^{cb}_{\mu}
-g^{2}\varepsilon^{ac}\varepsilon^{bd}A^{c}_{\mu}A^{d}_{\mu}\;,
\label{M}
\end{equation}
with the covariant derivative $D^{ab}_{\mu}$ defined as a
particular case of \eqref{covdev} by
\begin{equation}
D^{ab}_{\mu}=\d^{ab}\p_{\mu}-g\varepsilon^{ab}A_{\mu}\;.
\end{equation}
The inclusion of the horizon function \eqref{horizon} allows one
to implement the restriction of the domain of integration in the
Feynman path integral to the Gribov region, where the operator
\eqref{M} is strictly positive definite. As underlined in
\cite{Capri:2005tj,Capri:2006cz}, such a restriction is necessary
in order to deal with the Gribov copies.

\noindent In much the same way as the gauge invariant mass
operator \eqref{trmassop}, the horizon function \eqref{horizon}
also possesses a localized version, which can be obtained through
the introduction of a suitable pair of commuting auxiliary complex
fields $(\phi^{ab}_{\mu}, \bar{\phi}^{ab}_{\mu})$, and a pair of
anti-commuting ones $(\omega^{ab}_{\mu},\bar{\omega}^{ab}_{\mu})$
\cite{Capri:2006cz}
\begin{eqnarray}
S_{\mathrm{Horizon}}^{\mathrm{local}}&=&\int d^{4}\!x\,\Bigl\{
\bar\phi^{ab}_{\mu}{\cal
M}^{ac}\phi^{cb}_{\mu}-\bar\omega^{ab}_{\mu}{\cal
M}^{ac}\omega^{cb}_{\mu}+\bar\omega^{ab}_{\mu}{\cal
F}^{ac}\phi^{cb}_{\mu}+
\bar{M}^{ac}_{\mu\nu}\,D^{ab}_{\mu}\phi^{bc}_{\nu}+N^{ac}_{\mu\nu}
\,D^{ab}_{\mu}\bar\omega^{bc}_{\nu}+\nonumber\\
&&+\bar{N}^{ac}_{\mu\nu}\Bigl[D^{ab}_{\mu}\omega^{bc}_{\nu}+
g\varepsilon^{ab}(\partial_{\mu}c+
g\varepsilon^{de}A^{d}_{\mu}c^{e})\phi^{bc}_{\nu}\Bigr]+
M^{ac}_{\mu\nu}\Bigr[D^{ab}_{\mu}\bar\phi^{bc}_{\nu}+
g\varepsilon^{ab}(\partial_{\mu}c+
g\varepsilon^{de}A^{d}_{\mu}c^{e})\bar\omega^{bc}_{\nu}\Bigl]\Bigr\}\nonumber\\
&&+\alpha g^{2}\int d^{4}\!x\,\Bigl[\half
\left(\bar{\phi}^{ac}_{\mu}\phi^{ac}_{\mu}
-\bar{\omega}^{ac}_{\mu}\omega^{ac}_{\mu}\right)
\left(\bar{\phi}^{bd}_{\nu}\phi^{bd}_{\nu}
-\bar{\omega}^{bd}_{\nu}\omega^{bd}_{\nu}\right)
-\left(\bar{\phi}^{ac}_{\mu}\phi^{ac}_{\mu}
-\bar{\omega}^{ac}_{\mu}\omega^{ac}_{\mu}\right)\bar{c}^{b}c^{b}\nonumber\\
&&\phantom{+\alpha
g^{2}}+\bar{\omega}^{ac}_{\mu}\phi^{ac}_{\mu}\,b^{b}c^{b}
-\bar{\omega}^{ac}_{\mu}\phi^{ac}_{\mu}\,g\varepsilon^{bd}\bar{c}^{b}c^{d}c\Bigr]
\;, \label{hor_local}
\end{eqnarray}
with $\mathcal{F}^{ab}$ given by
\begin{eqnarray}
{\cal F}^{ab}&=&2g\varepsilon^{ac}(\partial_{\mu}c+
g\varepsilon^{de}A^{d}_{\mu}c^{e})D^{cb}_{\mu}+
g\varepsilon^{ab}\partial_{\mu}(\partial_{\mu}c+
g\varepsilon^{cd}A^{c}_{\mu}c^{d})+\nonumber\\
&&-g^{2}(\varepsilon^{ac}\varepsilon^{bd}+
\varepsilon^{ad}\varepsilon^{bc})A^{d}_{\mu}(D^{ce\!}_{\mu}c^{e}
+g\varepsilon^{ce}A^{e}_{\mu}c)\;,
\end{eqnarray}
and the sources $M^{ab}_{\mu\nu}$, $\bar{M}^{ab}_{\mu\nu}$,
$N^{ab}_{\mu\nu}$, and $\bar{N}^{ab}_{\mu\nu}$ are chosen in such a
way that their physical values must be taken as
\begin{eqnarray}
\bar{M}^{ab}_{\mu\nu}\Bigl|_{\mathrm{phys}}&=&
-M^{ab}_{\mu\nu}\Bigl|_{\mathrm{phys}}
=\delta^{ab}\delta_{\mu\nu}\gamma^{2}\;,\nonumber\\
\bar{N}^{ab}_{\mu\nu}\Bigl|_{\mathrm{phys}}&=&
N^{ab}_{\mu\nu}\Bigl|_{\mathrm{phys}}=0\;. \label{physvalues}
\end{eqnarray}
Notice that the last term in expression \eqref{hor_local}
introduces quartic interactions between the auxiliary fields,
being needed for renormalizability. Nevertheless, unlike the term
\eqref{lambda_term}, these quartic terms depend on the gauge
parameter $\alpha$ of \eqref{MAG},  which will be set to zero,
$\alpha\to0$, after the removal of the ultraviolet divergences.

\noindent Adding the local version of the horizon function
\eqref{hor_local} to the action \eqref{Sigma}, we obtain a new
starting  action which reads
\begin{equation}
S=S_{\mathrm{inv}}+S^{\mathrm{local}}_{\mathrm{horizon}}+S_{\lambda}+{\tilde S_{\mathrm{sources}}}
+{\tilde S_{\mathrm{ext}}}\;.\label{new_action}
\end{equation}
The first three terms of expression \eqref{new_action} are given
by \eqref{Sinv}, \eqref{hor_local}, and \eqref{lambda_term}
respectively, while the fourth term generalizes \eqref{sources},
including the new sources \eqref{physvalues}, namely
\begin{equation}
{\tilde S_{\mathrm{sources}}} = S_{\mathrm{sources}} + \int
d^{4}\!x\,\chi\Bigl(\bar{M}^{ab}_{\mu\nu}M^{ab}_{\mu\nu}
+\bar{N}^{ab}_{\mu\nu}N^{ab}_{\mu\nu}\Bigr)\;.
\end{equation}
Finally, the last term of \eqref{new_action} contains the coupling
of the external sources $ \Omega^{a}_{\mu},\tau^{a}_{\mu},
\xi^{a}_{\mu}, \Omega_{\mu}, L^{a}, L, \bar{Y}^{B}_{I},
{Y}^{B}_{I}, \bar{X}^{B}_{I}, X^{B}_{I}$ and $ \lambda^{ac}_{\mu},
\eta^{ae}_{\mu}, \rho^{ac}_{\mu}, \vartheta^{ae}_{\mu} $ to some
nonlinear operators needed for the BRST invariance of the model,
being given by
\begin{eqnarray}
{\tilde S_{\mathrm{ext}}}&\!\!\!=\!\!\!&\int
d^{4}\!x\,\biggl\{-\Omega^{a}_{\mu}\,D^{ab\!}_{\mu} c^{b}
-g\varepsilon^{ab}\tau^{a}_{\mu}A^{b}_{\mu}c
+\xi^{a}_{\mu}\Bigl[g\varepsilon^{ab}(D^{bc\!}_{\mu}c^{c})c
-\frac{g^{2}}{2}\varepsilon^{ab}\varepsilon^{cd}A^{b}_{\mu} c^{c}
c^{d}\Bigr]-\Omega_{\mu}(\partial_{\mu}c+
g\varepsilon^{ab}A^{a}_{\mu}c^{b})\nonumber\\
&&+g\varepsilon^{ab}L^{a}c^{b}c+
\frac{g}{2}\varepsilon^{ab}Lc^{a}c^{b}
+g\varepsilon^{ABC}c^{A}\left(\bar{Y}^{B}_{I}B^{C}_{I}
+{Y}^{B}_{I}\bar{B}^{C}_{I} -\bar{X}^{B}_{I}G^{C}_{I}
-{X}^{B}_{I}\bar{G}^{C}_{I}\right)
+g\varepsilon^{ab}\lambda^{ac}_{\mu}\phi^{bc}_{\mu}c\nonumber\\
&&+\eta^{ae}_{\mu}\Bigl[g\varepsilon^{ab}\omega^{be}_{\mu}c+
\frac{g^{2}}{2}\varepsilon^{ab}\varepsilon^{cd}\phi^{be}_{\mu}c^{c}c^{d}\Bigr]
+g\varepsilon^{ab}\rho^{ac}_{\mu}\bar\omega^{bc}_{\mu}c
-\vartheta^{ae}_{\mu}\Bigl[g\varepsilon^{ab}\bar\phi^{be}_{\mu}c
-\frac{g^{2}}{2}\varepsilon^{ab}\varepsilon^{cd}
\bar\omega^{be}_{\mu}c^{c}c^{d}\Bigr]\biggr\}\;.
\end{eqnarray}
The quantum numbers of the new fields and sources are displayed in
table \ref{table2}. \\\\Let us now proceed by giving the set of
Ward identities fulfilled by the action \eqref{new_action}. These
are:
\begin{itemize}
\item{The Slavnov-Taylor identity:
\begin{eqnarray}
\hspace{-20pt}{\cal S}(S)&\!\!\!\!\!=\!\!\!\!\!&\int
d^4\!x\,\biggl[\!\biggl(\frac{\delta S}{\delta\Omega^{a}_{\mu}}
+\frac{\delta S}{\delta\tau^{a}_{\mu}}\biggr)\!\frac{\delta
S}{\delta A^{a}_{\mu}} +\frac{\delta
S}{\delta\Omega_{\mu}}\frac{\delta S}{\delta A_{\mu}} +\frac{\delta
S}{\delta L^{A}}\frac{\delta S}{\delta c^{A}} +b^{A}\frac{\delta
S}{\delta\bar c^{A}}
+\!\left(\frac{\d{S}}{\d\bar{Y}^{A}_{I}}+G^{A}_{I}\right)\!\frac{\d{S}}{\d
B^{A}_{I}} +\frac{\d{S}}{\d
Y^{A}_{I}}\frac{\d\S}{\d\bar{B}^{A}_{I}}\nonumber\\
&&+\frac{\d{S}}{\d\bar{X}^{A}_{I}}\frac{\d{S}}{\d G^{A}_{I}}
+\left(\frac{\d{S}}{\d
X^{A}_{I}}+\bar{B}^{A}_{I}\right)\frac{\d{S}}{\d\bar{G}^{A}_{I}}
+\bar{V}_{I\mu\nu}\frac{\d{S}}{\d\bar{U}_{I\mu\nu}}
+U_{I\mu\nu}\frac{\d{S}}{\d V_{I\mu\nu}}
-\bar{Y}^{A}_{I}\frac{\d{S}}{\d\bar{X}^{A}_{I}}
+{X}^{A}_{I}\frac{\d{S}}{\d{Y}^{A}_{I}}\nonumber\\
&&+\omega^{ab}_{\mu}\frac{\delta
S}{\delta\phi^{ab}_{\mu}}+\bar\phi^{ab}_{\mu}\frac{\delta
S}{\delta\bar\omega^{ab}_{\mu}} +N^{ab}_{\mu\nu}\frac{\delta
S}{\delta M^{ab}_{\mu\nu}} -\bar{M}^{ab}_{\mu\nu}\frac{\delta
S}{\delta \bar{N}^{ab}_{\mu\nu}}
-(\Omega^{a}_{\mu}-\tau^{a}_{\mu})\frac{\delta
S}{\delta\xi^{a}_{\mu}} +\lambda^{ab}_{\mu}\frac{\delta
S}{\delta\eta^{ab}_{\mu}} +\rho^{ab}_{\mu}\frac{\delta
S}{\delta\vartheta^{ab}_{\mu}}\biggr]=0\;.
\end{eqnarray}}

\item{The global $U(8)$ invariance:
\begin{equation}
{\cal Q}^{ab}_{\mu\nu}(S)=0\;,
\end{equation}
with
\begin{eqnarray}
{\cal Q}^{ab}_{\mu\nu}&\equiv&\int
d^{4}\!x\,\biggl(\phi^{ca}_{\mu}\frac{\delta}{\delta\phi^{cb}_{\nu}}
-\bar\phi^{cb}_{\nu}\frac{\delta}{\delta\bar\phi^{ca}_{\mu}}+
\omega^{ca}_{\mu}\frac{\delta}{\delta\omega^{cb}_{\nu}}
-\bar\omega^{cb}_{\nu}\frac{\delta}{\delta\bar\omega^{ca}_{\mu}}
+M^{ca}_{\sigma\mu}\frac{\delta}{\delta M^{cb}_{\sigma\nu}}
-\bar{M}^{cb}_{\sigma\nu}\frac{\delta}{\delta
\bar{M}^{ca}_{\sigma\mu}}\nonumber\\
&&+N^{ca}_{\sigma\mu}\frac{\delta}{\delta N^{cb}_{\sigma\nu}}
-\bar{N}^{cb}_{\sigma\nu}\frac{\delta}{\delta
\bar{N}^{ca}_{\sigma\mu}}
+\vartheta^{ca}_{\mu}\frac{\d}{\d\vartheta^{cb}_{\nu}}
-\eta^{cb}_{\nu}\frac{\d}{\d\eta^{ca}_{\mu}}
+\rho^{ca}_{\mu}\frac{\d}{\d\rho^{cb}_{\nu}}
-\lambda^{cb}_{nu}\frac{\d}{\d\lambda^{ca}_{\mu}}\biggr)\;.\label{Q_8}
\end{eqnarray}
The presence of this global invariance $U(8)$ allows us to make
use of a composite index $i\equiv(a,\mu)$, with $i=1,\dots,8$.
Thus, from now on, we set
\begin{eqnarray}
(\phi^{ab}_{\mu},\bar\phi^{ab}_{\mu},\omega^{ab}_{\mu},
\bar\omega^{ab}_{\mu})&=&
(\phi^{a}_{i},\bar\phi^{a}_{i},\omega^{a}_{i},
\bar\omega^{a}_{i})\;,\\
(M^{ab}_{\mu\nu},\bar{M}^{ab}_{\mu\nu},N^{ab}_{\mu\nu},\bar{N}^{ab}_{\mu\nu})&=&
(M^{a}_{\mu i},\bar{M}^{a}_{\mu i},N^{a}_{\mu i},\bar{N}^{a}_{\mu
i})\;,\\
(\vartheta^{ab}_{\mu},\eta^{ab}_{\mu},\rho^{ab}_{\mu},\lambda^{ab}_{\mu})
&=&(\vartheta^{a}_{i},\eta^{a}_{i},\rho^{a}_{i},\lambda^{a}_{i})\;.
\end{eqnarray}
The trace of \eqref{Q_8} defines a new
$\mathcal{Q}_{8}$\emph{-charge} whose nonvanishing values are
displayed in table \ref{table2}.}

\begin{table}[t] \centering
{\small\begin{tabular}{lccccccccccccccc} \hline\hline
&$\phi$&$\bar{\phi}$&$\omega$&$\bar{\omega}$&$M$&$\bar{M}$&$N$&$\bar{N}$&$\vartheta$
&$\eta$&
$\rho$&$\lambda$&$\Omega$&$\tau$&$\xi\!\phantom{\Bigl|}$\\
\hline
dimension&$1$&$1$&$1$&$1$&$2$&$2$&$2$&$2$&$3$&$3$&$3$&$3$&$3$&$3$&$3$\\
gh.
number&$0$&$0$&$1$&$-1$&$0$&$0$&$1$&$-1$&$-1$&$-2$&$0$&$-1$&$-1$&$-1$&$-2$\\
$\mathcal{Q}_{8}$-charge&$1$&$-1$&$1$&$-1$&$1$&$-1$&$1$&$-1$&$1$
&$-1$&$1$&$-1$&$0$&$0$&$0$\\
\hline\hline
\end{tabular}}
\caption{Quantum numbers of the fields and sources} \label{table2}
\end{table}

\item{Symmetries involving the Faddeev-Popov ghost fields and the
 localizing fields:
\begin{eqnarray}
\mathcal{W}^{(1)}_{i}(S)&=&\int d^4\!x\,\biggl[\phi^{a}_{i}\frac{\d
S}{\d\bar{c}^{a}} +c^{a}\frac{\d S}{\delta\bar{\phi}^{a}_{i}}
+M^{a}_{\mu i}\frac{\delta S}{\delta\Omega^{a}_{\mu}}
-\vartheta^{a}_{i}\frac{\delta S}{\delta L^{a}} +\frac{\delta
S}{\delta\lambda^{a}_{i}} \frac{\delta S}{\delta
b^{a}}\biggr]=0\;,\label{w1}\\
\mathcal{W}^{(2)}_{i}(S)&=&\int
d^4\!x\,\biggl[\bar\omega^{a}_{i}\frac{\delta S}{\delta\bar
c^{a}}-c^{a}\frac{\delta S}{\delta\omega^{a}_{i}}-\bar{N}^{a}_{\mu i
}\frac{\delta S}{\delta\Omega^{a}_{\mu}} +\eta^{a}_{i} \frac{\delta
S}{\delta L^{a}} +\frac{\delta S}{\delta\rho^{a}_{i}} \frac{\delta
S}{\delta
b^{a}}\biggr]=0\;,\label{w2}\\
\mathcal{W}^{(3)}_{i}(S)&=&\int d^4\!x\,\biggl[\biggl(\frac{\delta
S}{\delta\lambda^{a}_{i}}+ \omega^{a}_{i}\biggr)\frac{\delta
S}{\delta\bar c^{a}} +\frac{\delta
S}{\delta\eta^{a}_{i}}\frac{\delta S}{\delta b^{a}}
+\biggl(\frac{\delta S}{\delta\bar\phi^{a}_{i}}
-\rho^{a}_{i}\biggr)\frac{\delta S}{\delta L^{a}}\nonumber\\
&&+c^{a}\frac{\delta S}{\delta\bar\omega^{a}_{i}} -M^{a}_{\mu
i}\frac{\delta S}{\delta\xi^{a}_{\mu}} +N^{a}_{\mu i}\frac{\delta
S}{\delta\Omega^{a}_\mu}\biggr]=0\;,
\label{w3}\\
\mathcal{W}^{(4)}_{i}(S)&=&\int d^4\!x\,\biggl[\biggl(\frac{\delta
S}{\delta\rho^{a}_{i}} -\bar\phi^{a}_{i}\biggr)\frac{\delta
S}{\delta\bar c^{a}} +\frac{\delta S}{\delta\vartheta^{a}_{i}}
\frac{\delta S}{\delta b^{a}} +\biggl(\frac{\delta
S}{\delta\omega^{a}_{i}}
-\lambda^{ab}_{\nu}\biggr)\frac{\delta S}{\delta L^{a}}\nonumber\\
&&-c^{a}\frac{\delta S}{\delta\phi^{a}_{i}}+\bar{N}^{a}_{\mu i
}\frac{\delta S}{\delta\xi^{a}_{\mu}}-\bar{M}^{a}_{\mu
i}\frac{\delta S}{\delta\Omega^{a}_\mu}\biggr]=0\;. \label{w4}
\end{eqnarray}}

\item{The exact rigid symmetries associated to the horizon
function:
\begin{eqnarray}
R^{(1)}_{ij}(S)&=&\int d^{4}\!x\,\biggl(\phi^{a}_{i}\frac{\delta
S}{\delta\omega^{a}_{j}} -\bar\omega^{a}_{j}\frac{\delta
S}{\delta\bar\phi^{a}_{i}}+M^{a}_{\mu i}\frac{\delta S}{\delta
N^{a}_{\mu j}}+\bar{N}^{a}_{\mu j}\frac{\delta S}{\delta
\bar{M}^{a}_{\mu i}}+\vartheta^{a}_{i}\frac{\delta
S}{\delta\rho^{a}_{j}} -\eta^{a}_{j}\frac{\delta
S}{\delta\lambda^{a}_{i}}\biggr)=0\;,
\label{R1}\\\nonumber\\
R^{(2)}(S)&=&\int d^{4}\!x\,\biggl(\bar\omega^{a}_{i} \frac{\delta
S}{\delta\omega^{a}_{i}}-\bar{N}^{a}_{\mu i} \frac{\delta S}{\delta
N^{a}_{\mu i}}-\eta^{a}_{i}\frac{\delta
S}{\delta\rho^{a}_{i}}\biggr)=0\;,
\label{R2}\\\nonumber\\
R^{(3)}(S)&=&\int d^{4}\!x\,\biggl(
\bar\omega^{a}_{i}\frac{\delta{S}}{\delta\phi^{a}_{i}}
-\bar\phi^{a}_{i}\frac{\delta{S}}{\delta\omega^{a}_{i}} -
\bar{N}^{a}_{\mu i}\frac{\delta{S}}{\delta M^{a}_{\mu
i}}-\bar{M}^{a}_{\mu i}\frac{\delta{S}}{\delta N^{a}_{\mu
i}}-\eta^{a}_{i}\frac{\delta{S}}{\delta\vartheta^{a}_{i}}+
\lambda^{a}_{i}\frac{\delta{S}}{\delta\rho^{a}_{i}}\biggr)=0\;.
\label{R3}
\end{eqnarray}}

\item{The global $U(6)$ invariance:
\begin{equation}
\mathcal{Q}_{IJ}(S)=0\;,
\end{equation}
with the operator $\mathcal{Q}_{IJ}$ given by \eqref{Q_op}.}

\item{The exact rigid symmetries associated to the mass operator:
\begin{equation}
\mathcal{R}^{(\mathcal{N})}_{IJ}(S)=0\;,
\end{equation}
with the same operators $\mathcal{R}^{(\mathcal{N})}_{IJ}$,
$\mathcal{N}=1,2,3,4$, already defined in \eqref{rigit_ops}.}

\item{The diagonal $U(1)$ Ward identity:
\begin{equation}
{\cal W}^{3}(S)=-\partial^{2}b\;,
\end{equation}
with
\begin{eqnarray}
{\cal W}^{3}&\equiv&\partial_{\mu}\frac{\delta}{\delta
A_{\mu}}+g\varepsilon^{ab}\biggl(A^{a}_{\mu} \frac{\delta}{\delta
A^{b}_{\mu}}+b^{a}\frac{\delta}{\delta
b^{b}}+c^{a}\frac{\delta}{\delta c^{b}}+ \bar
c^{a}\frac{\delta}{\delta \bar c^{b}}+
\phi^{a}_{i}\frac{\delta}{\delta \phi^{b}_{i}}+
\bar\phi^{a}_{i}\frac{\delta}{\delta \bar\phi^{b}_{i}}\nonumber\\
&&+\omega^{a}_{i}\frac{\delta}{\delta \omega^{b}_{i}}+
\bar\omega^{a}_{i}\frac{\delta}{\delta \bar\omega^{b}_{i}}
+\Omega^{a}_{\mu}\frac{\delta}{\delta \Omega^{b}_{\mu}}+
\tau^{a}_{\mu}\frac{\delta}{\delta \tau^{b}_{\mu}}+
\xi^{a}_{\mu}\frac{\delta}{\delta \xi^{b}_{\mu}}+ M^{a}_{\mu
i}\frac{\delta}{\delta M^{b}_{\mu i}}+ \bar{M}^{a}_{\mu
i}\frac{\delta}{\delta \bar{M}^{b}_{\mu
 i}}\nonumber\\
&&+N^{a}_{\mu i}\frac{\delta}{\delta N^{b}_{\mu i}}+
\bar{N}^{a}_{\mu i}\frac{\delta}{\delta \bar{N}^{b}_{\mu i}}+
\eta^{a}_{i}\frac{\delta}{\delta \eta^{b}_{i}}+
\vartheta^{a}_{i}\frac{\delta}{\delta \vartheta^{b}_{i}}+
\lambda^{a}_{i}\frac{\delta}{\delta \lambda^{b}_{i}}+
\rho^{a}_{i}\frac{\delta}{\delta \rho^{b}_{i}}+
L^{a}\frac{\delta}{\delta L^{b}}\nonumber\\
&&+B^{a}_{I}\frac{\d}{\d B^{b}_{I}} +\bar{B}^{a}_{I}\frac{\d}{\d
\bar{B}^{b}_{I}} +G^{a}_{I}\frac{\d}{\d G^{b}_{I}}
+\bar{G}^{a}_{I}\frac{\d}{\d \bar{G}^{b}_{I}} +Y^{a}_{I}\frac{\d}{\d
Y^{b}_{I}} +\bar{Y}^{a}_{I}\frac{\d}{\d \bar{Y}^{b}_{I}}
+X^{a}_{I}\frac{\d}{\d X^{b}_{I}} +\bar{X}^{a}_{I}\frac{\d}{\d
\bar{X}^{b}_{I}}\biggr)\;.
\end{eqnarray}}

\item{The off-diagonal SL(2,$\mathbb{R}$) identity:
\begin{equation}
\mathcal{D}(S)=\;\Int\,\left(c^{a}\frac{\d S}{\d\bar{c}^{a}}
+\frac{\d S}{\d L^{a}}\frac{\d S}{\d b^{a}}\right)=0\;.
\end{equation}}

\item{The diagonal gauge fixing:
\begin{equation}
\frac{\d S}{\d b}=\partial_{\mu}A_{\mu}\;.
\end{equation}}

\item{The diagonal anti-ghost equation:
\begin{equation}
\frac{\d S}{\d\bar{c}}+\p_{\mu}\frac{\d S}{\d\Omega_{\mu}}=0\;.
\end{equation}}
\end{itemize}

\noindent We notice here that no diagonal ghost equation, similar
to \eqref{diag_gh}, holds when the horizon function is taken into
account \cite{Capri:2006cz}. However, the set of Ward identities
listed above forbids the presence of  counterterms like
\begin{equation}
\int d^{4}\!x\,
(\bar{\phi}^{a}_{i}\phi^{a}_{i}-\bar{\omega}^{a}_{i}\omega^{a}_{i})
(\bar{B}^{A}_{I}B^{A}_{I}-\bar{G}^{A}_{I}G^{A}_{I})\;,
\end{equation}
as well as of any other counterterm which would mix fields
associated with the two different nonlocal operators
\eqref{trmassop} and \eqref{horizon}. Therefore, in complete
analogy with the case of the Landau gauge \cite{Capri:2007ix}, the
two operators \eqref{trmassop} and \eqref{horizon} do not mix, due
to the rich symmetry content of the resulting local action.
Moreover, it turns out that the most general allowed counterterm
can be in fact reabsorbed in the starting action,
\eqref{new_action}, through a redefinition of fields, parameters
and sources, ensuring the renormalizability of the mass operator
\eqref{trmassop} in the presence of the horizon function
\eqref{horizon}.

\end{document}